\documentclass[12pt]{article}
\usepackage{color}
\usepackage{authblk}
\usepackage{hyperref} 
\usepackage{hanging}

\let\ssection=\section
\renewcommand{\section}{\setcounter{equation}{0}\ssection}

\newcommand\mathC{\mkern1mu\raise2.2pt\hbox{$\scriptscriptstyle|$}
        {\mkern-7mu\rm C}}

\newcommand\be{\begin{equation}}
\newcommand\ee{\end{equation}}

\title{{\bf\large{Scientific Realism and Primordial Cosmology}}}

\author[1,2]{Feraz Azhar \thanks{Email: fa232@cam.ac.uk}}
\affil[1]{Department of History and Philosophy of Science, University of Cambridge, Free School Lane, Cambridge CB2 3RH, United Kingdom}
\affil[2]{Program in Science, Technology, and Society, Massachusetts Institute of Technology, Cambridge, MA 02139, United States of America}

\author[3]{Jeremy Butterfield \thanks{Email: jb56@cam.ac.uk}}
\affil[3]{Trinity College, Cambridge, CB2 1TQ, Cambridge, United Kingdom}

\date{}
\begin{document}

\catcode`\'=\active\relax\def'{^{\prime}}\catcode`\'=12\relax

\maketitle

\vspace{-1cm}
\begin{center}
An abridged version will appear in {\em The Routledge Handbook of Scientific Realism}, ed. Juha Saatsi.
\end{center}

\begin{center}
(Dated: Monday 13 June 2016)
\end{center}

\begin{abstract}
We discuss scientific realism from the perspective of modern cosmology, especially primordial cosmology: i.e. the cosmological investigation of the very early universe. 

We first (Section \ref{gen}) state our allegiance to scientific realism, and discuss what insights about it cosmology might yield, as against ``just'' supplying scientific claims that philosophers can then evaluate.  In particular, we discuss: the idea of laws of cosmology, and limitations on ascertaining the global structure of spacetime. Then we review some of what is now known about the early universe (Section \ref{know}):  meaning, roughly, from a thousandth of a second after the Big Bang onwards(!).

The rest of the paper takes up two issues about primordial cosmology, i.e. the {\em very} early universe, where ``very early'' means, roughly, much earlier (logarithmically) than one second after the Big Bang: say, less than $10^{-11}$ seconds. Both issues illustrate that familiar philosophical threat to scientific realism, the under-determination of theory by data---on a cosmic scale. 

The first issue  (Section \ref{plan}) concerns the difficulty of observationally probing the very early universe. More specifically, the difficulty is to ascertain details of the putative inflationary epoch. The second issue (Section \ref{prob}) concerns difficulties about confirming a cosmological theory that postulates a multiverse, i.e. a set of domains (universes) each of whose inhabitants (if any) cannot directly observe, or otherwise causally interact with, other domains. This again concerns inflation, since many inflationary models postulate a multiverse.
 
For all these issues, it will be clear that much remains unsettled, as regards both physics and philosophy. But we will maintain that these remaining controversies do not threaten scientific realism. \\

{\bf Keywords}: scientific realism, primordial cosmology, inflation, multiverse

\end{abstract}

\newpage
\tableofcontents

\newpage

\section{Introduction}\label{intro}
We will discuss scientific realism from the perspective of cosmology, especially primordial cosmology: i.e. the cosmological investigation of the very early universe.  

We first (Section \ref{gen}) state our allegiance to scientific realism, and discuss what insights about it cosmology might yield, as against ``just'' supplying scientific claims that philosophers can then evaluate. In particular, we set aside (Section \ref{laws?}) the traditional methodological worry that cosmology cannot be a science; (because, it is alleged, there being only one universe means that any putative laws of cosmology would have only one instance). We also discuss limitations on ascertaining the global structure of any universe described by general relativity (Section \ref{manchak}). Then in Section \ref{know}, we review in a realist spirit, some of what is now known about the early universe. Here, ``early'' will mean, for us, times later than about $10^{-11}$ seconds after the Big Bang (!).

The rest of the paper addresses, in more detail, two issues in cosmology that bear on scientific realism: especially the theme, familiar in philosophy of science, of the under-determination of theory by data. But the issues do not (we believe!) threaten scientific realism. Rather, they clarify---and agreed: limit---what a  scientific realist should take as definitively established by modern cosmology. They both concern {\em primordial} cosmology; where ``primordial'',  and similarly ``very early'',  means for us: ``so early that the energies are higher than those which our established theories successfully describe''. This will turn out to mean: times earlier than about $10^{-11}$ seconds after the Big Bang. The two issues each have a vast literature, written by cosmologists: who have proven---we are happy to report---to be very insightful about the conceptual, indeed philosophical, issues involved.  

The first issue (Section \ref{plan}) concerns the difficulty of observationally probing the very early universe. Thus expressed, this is hardly news: we would expect it to be difficult! But the issue is more specific.  In the last thirty years, it has become widely accepted that at times much earlier (logarithmically) than one second after the Big Bang, there was an epoch of accelerating expansion (dubbed {\em inflation}) in which the universe grew by many orders of magnitude. The conjectured mechanism for this expansion was a physical field, the {\em inflaton} field, $\phi$, subject to an appropriate potential $V(\phi)$; (or maybe by a set of such fields, but most models use a single field). The evidence for this inflationary framework lies principally in (i) its solving three problems that beset previous general relativistic cosmological models (the flatness, horizon and monopole problems), and (ii) its explanation of some main features of the cosmic microwave background radiation (CMB).  However, this evidence leaves very {\em undetermined} the detailed physics of the inflationary epoch: in particular, it allows many choices for the shape of the potential $V(\phi)$, and there are nowadays many different models of inflation.

 The second issue (Section \ref{prob}) concerns the difficulty of confirming a cosmological theory  that postulates a {\em multiverse}, i.e. a set of domains (universes) each of whose inhabitants (if any) cannot directly observe, or otherwise causally interact with, other domains. This issue arises because many models of inflation amount to such a theory. That is: according to many such models, in an epoch even earlier than that addressed in our first issue---and even harder to access---a vast number of different domains came into being, of which just one  gave rise to the universe we see around us, while the others  ``branched off'', and are forever inaccessible to us. That is, of course, rough speaking: more details in Section \ref{prob}. For now, we emphasize that this picture is very speculative: the physics describing this earlier epoch is not known. The energies and temperatures involved are far above those that are so successfully described by the standard model of elementary particle physics, and far above those probed by the observations described in relation to our first issue (as in Section \ref{plan}). 
 
Nevertheless, cosmologists have addressed the methodological question how we can possibly confirm a multiverse theory, often by using toy models of such a multiverse. The difficulty is not just that it is hard to get evidence. We also need to allow for the fact that what evidence we get may be greatly influenced, indeed distorted, by our means of observation: that is, by our location in the multiverse.  Thus one main, and well-known, aspect concerns the legitimacy of anthropic explanations. That is, in broad terms: we need to ask: How satisfactory is it to explain an observed fact by appealing to its absence being incompatible with the existence of an observer?'
     
For all the issues we discuss, it will be clear that much remains unsettled, as regards both physics and philosophy. But we will maintain that these remaining controversies do not threaten scientific realism.  We can already state the main reason why not. In Section \ref{gen}, we will take scientific realism to be a claim along the lines ``we can know, indeed do know, about the unobservable''. But that does not imply that---and it is simply no part of scientific realism to claim that---``all the unobservable is known''; or even, ``all the unobservable is knowable''. For example, we will maintain that the scientific realist can  perfectly well accept---should accept!---that:\\
 \indent (i) the global structure of space-time may be unknowable (indeed: will provably be unknowable, if general relativity is true); \\
 \indent (ii) dark matter and dark energy are known to exist, but that their nature is unknown; \\
 \indent (iii) even assuming there is an inflaton field $\phi$, the potential $V(\phi)$ governing it may be unknowable; and \\
 \indent  (iv) a theory that postulates a multiverse faces special difficulties about confirmation.\\
 All these four admissions, corresponding to Sections  \ref{manchak}, \ref{know}, \ref{plan} and \ref{prob} respectively, are just `epistemic modesty'. They are compatible with the characteristic `epistemic optimism' of scientific realism---that much is already known, and  yet more can be: an upbeat note on which our final summary (Section \ref{concl}) will end.\\
 \newpage
 \emph{Issues that we set aside}:---\\
We should also register at the outset several other ways in which our discussion will leave issues unsettled. Partly this is a matter of setting issues aside just because we intend to write a review of mainstream ideas in a limited space---and with limited expertise! And partly it is a matter of the issues being open.

First: there are broad issues directly about scientific realism. For example:\\
\indent (A): Does the fact that in cosmology, and indeed astronomy, we cannot manipulate the objects and events in question as we do in other sciences, undercut claims of scientific realism; (or of its variants like Hacking's entity realism (Hacking 1983, Miller 2016)? \\
\indent (B): Does this fact undercut claims about causation, at least when understood in terms of manipulations or interventions (Woodward 2003)?\\
\indent (C): Does the difficulty of observationally probing the early universe undercut science's usual strategy of confirming new theoretical postulates by finding independent lines of access to the postulated entities? (This of course relates directly to our chosen focus on the under-determination of theory by data.)\\
Such issues, especially the last, have been discussed very judiciously, and with more detail than here, by Smeenk (2013: Sections 6--8; 2014: Sections 6--8; 2016: Sections 3--4).

Second: there are no less than five specific issues on which this essay might have focused (rather than the under-determination of theory by data)---all of which bear on scientific realism. We list them in a roughly increasing order of specificity vis-\`{a}-vis primordial cosmology. Though we will touch on some of them later, we will mostly set them aside; so here we also give some references.

(1): In recent decades, developments in high-energy physics, such as string theory and the recognition that most theories are effective, i.e. limited in their range of validity, has made problematic the confirmation of putatively fundamental theories: and thereby also, the defence of scientific realism. Cf. Dawid (this volume, 2013).

(2): At sufficiently early times after the Big Bang, energies are so high that atoms and even nuclei `melt', so that the proverbial clock and rod with which we measure time and space cannot possibly exist. So we need to scrutinize the limits of application of our temporal and spatial concepts in such regimes. This scrutiny has been undertaken by Rugh and Zinkernagel in several papers (2009, 2016).

(3): Modern cosmology, especially inflationary cosmology, makes much use of probability measures over, for example, some set of possible initial states in a cosmological model. How to define these measures rigorously, and how to then justify the choice of one measure rather than another, are often hard, and disputed, questions. Cf. for example, Schiffrin and Wald (2012); and for  philosophers' views, Koperski (2005), Norton (2010) and Curiel (2015).

(4): Inflationary cosmology proposes that quantum fluctuations in the inflaton field became classical and generated slight variations in matter density at early times: variations that led to slight anisotropies in the cosmic microwave background radiation (CMB), which were then magnified by matter clumping together under gravity, leading to stars and galaxies. We give some details of this remarkable mechanism in Section 4.2.2. But we should note at the outset that in the transition from quantum fluctuations to classical fluctuations, one faces quantum theory's notorious measurement problem! After all, `quantum fluctuation' really means `non-zero amplitude for more than one alternative' while `classical fluctuation' means (less puzzlingly!) `jitter in the actual possessed value of a given variable'. Of course, practitioners of inflationary cosmology recognize this; and many appeal to decoherence as a solution. But we believe, along with most {\em aficionados} of the measurement problem, that decoherence, though important, is not a complete solution. So the issue remains open; and fortunately, some foundationally-inclined cosmologists pursue it. Cf. for example, Perez et al. (2006), Sudarsky (2011), Ca\~{n}ate et al. (2013), Colin and Valentini (2016). 

(5): The hypothesis of a cosmological multiverse raises several issues of scientific method additional to those we will address in Section 5. One is whether the difficulty of confirming the hypothesis, and-or the ensuing need to accept anthropic explanations, prompt a revision in our conception of scientific explanation, or more generally in our conception of science and its method. This has been the subject of considerable debate: not surprisingly, since it borders on general questions about the aim and scope of science, and the perhaps special role of cosmology in humankind's search to understand the universe and our place in it. Surveys of these issues can be found in, for example, Carr (2014, especially Sections 4--6) and Ellis (2014, Sections 6, 8; 2016, especially Sections 2, 3, 6). Ellis and Silk (2014) is a good example of skepticism about the multiverse.   

So much by way of a list of issues to be set aside. (The list format is of course not meant to deny that the issues are connected. They obviously are: for example, conceptual advances about fundamental theories, under (1) or (4), might help with (3)'s problems about measures.) We now turn to what we have promised we will address \dots

\section{Cosmology as a special case for scientific realism}\label{gen}
We will first state our allegiance to scientific realism, and discuss what insights about it cosmology might yield, as against ``just'' supplying scientific claims that philosophers can then evaluate (Section \ref{bear}). Then we set aside the traditional methodological worry that cosmology cannot be a science, because, it is alleged, there cannot be laws of cosmology (Section \ref{laws?}). Finally, we review some limitations on ascertaining the structure of any universe described by general relativity (Section  \ref{manchak}).

\subsection{Scientific realism, and how cosmology bears on it}\label{bear}
We take scientific realism to be the doctrine that most of the statements of the mature scientific theories that we accept are true, or approximately true, whether the statement is about observable or unobservable states of affairs.  Here, ``true'' is to be understood in a straightforward correspondence sense, as given by classical  referential semantics. And, accordingly, scientific realism holds our acceptance of these theories to involve believing (most of) these statements---i.e. believing them to be true in a straightforward correspondence sense. This characterization goes back, of course, to van Fraassen (1980: 7--9). It is not the only characterization: judicious discussions of alternatives include Stanford (2006: 3--25, 141f.) and Chakravartty (2011: Section 1). But it is a widely adopted characterization---and will do for our purposes. 

We are scientific realists in this sense. We concede that to defend this position in general requires precision about the vague words ``mature'',  ``accept'', ``observable'', ``true'' and (perhaps especially) ``approximately true''. But we will leave this general defense to more competent philosophers (the other Chapters in this volume, and e.g. Psillos (1999, 2009)). Our theme is, rather, the relations between scientific realism and cosmology; and our main claims about them, in particular that cosmology gives no trouble to scientific realism, will not need such precision.

At first sight, this theme can seem unpromising, or at least limited. For scientific realism is a general philosophical doctrine; so one naturally expects to assess it by distinctively philosophical arguments.  And at first sight, this suggests that cosmology, or indeed any science, cannot be expected to help in the assessment: a science's  results, theoretical and observational, can hardly be expected to determine what our attitude to these very results (or other scientific results) should be. So it can seem that our theme is merely a matter of  cosmology providing examples of scientific claims and theories that illustrate the philosophical theses of scientific realists---or perhaps, of their various opponents: in short, a matter of cosmology providing case-studies for philosophy of science. 

We think there are two main replies to this skepticism. In short, they are as follows. (1): Cosmology  providing case-studies can be a rich theme, not a limited one. (2): The distinction between the philosophical and the empirical is not as sharp and straightforward as this skepticism assumes; and in fact cosmology raises various issues that bear on scientific realism---not by cosmology's results straightforwardly illustrating it (or threatening it), but by cosmology prompting a general philosophical question (or a whole line of thought) about it. We shall first say a little about (1); and then turn to (2)---to which most of the paper is devoted.   

(1): Obviously, a case-study can be rich in its philosophical morals. But  modern cosmology, with its truly stupendous knowledge claims, prompts a more specific point. Cosmologists nowadays claim to have established, for example, that a second after the Big Bang, the entire material contents of the universe we now see were confined in a dense fireball, with a temperature of about $10^{10}$ K and a density of about $2000$ kilograms per cubic centimeter. Hearing this, surely every philosopher (whether a scientific realist or not!) feels a school-child's thrill---quickly followed by worrying how we could ever know such a proposition? 

Our own view is  that the cosmologists are {\em right}. That is: this particular claim, and countless other  claims about the overall history of the universe from about the one-second epoch onwards, are now, and will forever remain, as well established as countless other scientific facts, e.g. that plants photosynthesize and that insulin has fifty-one amino acids. 

In Section \ref{know}, we will briefly defend this scientific realist view of the results of modern cosmology. But we admit  that the general task of assessing how scientific realism fares in today's cosmological theories would be very ambitious: by no means, an unpromising or limited endeavor. Indeed, it would be too ambitious for us: we will duck out of a general defence of our form of scientific realism for today's cosmological theories. For it would involve two major projects. One would have to first define what parts of these theories are indeed ``mature and accepted''. Then one would have to argue that most of these parts are ``true, or approximately true'' in a correspondence sense, whether they are about observable or unobservable states of affairs.  These projects outstrip both our knowledge of cosmology, and (a rather different matter) our knowledge of what is the state of play in cosmology: i.e. what the community of cosmologists regards as accepted. 

To illustrate the substantial questions that arise here, we mention the obvious topics: dark matter, and dark energy. (Ruiz-Lapuente (2010) is a fine collection, including both observational and theoretical perspectives; Massimi and Peacock (2015) is an introduction for philosophers.) We would say---along with most cosmologists, of course---that:\\
\indent (i) both of these are known to exist, and indeed known to dominate the matter-energy content of the universe; but \\
\indent (ii) they are not observable (at least: not yet!), and their nature is wholly unknown. \\
Both (i) and (ii) obviously raise questions of both physics and philosophy. Thus: in what sense are dark matter and dark energy known to exist, yet unobservable  (at least:  not yet ``directly observable'')? And does our present evidence for them warrant taking our present statements about them to be true, in a correspondence sense---even while we admit, {\em \`{a} la} (ii), that their nature is wholly unknown?

(2): But even without undertaking (1)'s ambitious task of assessing how scientific realism fares in today's cosmological theories, there is much to discuss under the theme of ``scientific realism and cosmology''. For the distinction between the philosophical and the empirical is not crisp or straightforward---as is obvious from all modern philosophy of science. (Here it is usual to cite Quine (1953). But we maintain that Quine was unfair to Carnap, and that there are much more nuanced treatments of the relation between the philosophical and the empirical, and between the analytic and the synthetic; cf. Putnam (1962), Stein (1992, 1994).) 

And indeed, there are various issues in cosmology, that---rather than straightforwardly supporting or threatening scientific realism---instead prompt a question (or a whole theme) about it.  We will first (Section \ref{laws?}) set aside one such theme, about the role of laws in cosmology, since we think that nowadays it is a non-issue---as do most cosmologists, and philosophers of cosmology. Then we will review some theorems in general relativity suggesting we cannot know the global structure of spacetime (Section \ref{manchak}). As we will see, this gives a different  perspective on the familiar philosophical theme of under-determination of theory by data, viz. by emphasizing the need, not just to obtain data, but to gather it together. Then in Section 3f., we will see how cosmology threatens the usual philosophical distinction between (i) under-determination by all data one could in principle obtain, and (ii) under-determination by all data obtainable in practice, or up to a certain stage of enquiry. (Following Sklar (1975: 380f.), (ii) is often called `transient under-determination'.) For data about the early universe is so hard to get that what is not obtainable in practice looks very much unobtainable in principle!

\subsection{No laws of cosmology? No worries}\label{laws?}
There is an obvious ``one-liner'' objection one might raise about the idea that cosmology is a science: as follows. Since there is, by definition,  only one universe,  any putative laws of cosmology could have only one instance.\footnote{Here and in the rest of this Subsection, ``universe'' can be taken as broadly as possible, so that the discussion also covers a cosmology that posits a multiverse: i.e. for such a cosmology, please read ``universe'' as ``multiverse''.} But laws are usually taken to be, or at least to imply, suitable true generalizations: where what counts as `suitable' is disputed, but is usually taken to imply  having many more instances than just one!\footnote{As recounted in countless discussions of laws of nature: the dispute mostly concerns distinguishing laws from ``accidents'', i.e. merely accidentally true generalizations; and one tempting way to make the distinction is to say that in a law, the universal quantifier ``all'' is unrestricted in scope---which is meant to make for having many instances.} So if a science aims to formulate laws, it seems that cosmology cannot be a science. 

The answer to this objection is clear. It is that cosmology can perfectly well be a science, without formulating, or aiming to formulate, even one law whose instances are universes. We say this, not because (as some philosophers maintain, e.g. Giere (1999))  science in general does not need to formulate laws, and need only formulate models: but because the laws of cosmology can just be  the laws of the various physical theories  that describe parts of the universe. Thus there is an ambiguity in a phrase like ``a law governing a universe'', and similarly, in claims like ``cosmology formulates laws governing universes''.  This can mean: either ``a law whose instances are universes'',  in which case we  reject the claim; or  ``a law whose instances are (as in other sciences) parts of the universe (i.e. objects, events, states of affairs), but which is called ``a law governing a universe'' because it is important in describing and-or explaining the spatially and temporally very large-scale features of the universe, which are the business of cosmology''---in which case we accept the claim. In short: the fact that cosmology has the whole universe as its subject-matter---since it aims to describe and explain the very large-scale features of the whole universe---is perfectly compatible with its laws being the laws of the various {\em local} physical theories: such as, in present-day cosmology, the laws of the various quantum field theories, the various theories of statistical mechanics and hydrodynamics, and special and general relativity. 

There is a good analogy here between cosmology and geology. Both have a special subject-matter---the universe and the Earth, respectively---about which they aim to describe and explain, not every feature, but certain large-scale features. To do this, they only need laws governing the parts of their subject-matter relevant to their descriptive and explanatory aims: they do not need laws whose {\em instances} are universes, or Earths. Thus cosmology is no more impugned as a science by the existence of only one universe, than geology is by the existence of only one planet Earth (Cleland 2002, Butterfield 2012: 4--7). (Agreed: with the discovery of exoplanets, `geology' might come to mean the science of all planets, or of all Earth-like planets. But the analogy remains good, with the current meaning of `geology'.) 

We said that this answer to the objection is clear. But we should note that some sixty years ago, the objection was actively discussed. This was because it was entangled with another more specific debate, about whether laws could describe a putative origin of the universe ({\em cosmogony}); and this debate was vivid, and involved cosmologists, because it related to the rival claims of the Big Bang and steady-state theories---the latter of course denying that there was such an origin. Thus if you held both that laws could not describe an origin of the universe, and that science aimed to formulate laws, then you would be minded to favor the steady-state theory, since it side-stepped this apparent limitation on science. 

Besides, this debate involved other questions, which again relate to the nature and role of laws in cosmology. One question, more philosophical than physical, concerned cosmogony in general: is the idea of an origin of the universe, i.e. a beginning of time, coherent? This had of course long been struggled with by philosophers, including Aristotle and Kant. A second question was closer to physics: how if at all could we justify the powerful, simplifying symmetry principles that were imposed on our cosmological models?  This second question was closely related to the Big Bang vs. steady-state debate, since it took two more specific forms, one for each side of the debate. Namely:\\
\indent (1): How could an  advocate of the Big Bang justify the cosmological principle (CP) of the Big Bang (Friedmann-Robertson-Walker: FRW) models?  (The CP requires that on sufficiently large length scales, the universe is spatially isotropic and homogeneous.) And: \\
\indent (2): How could a steady-state theorist justify strengthening the CP's requirements, by imposing the perfect cosmological principle (PCP): which added to spatial isotropy and homogeneity, the requirement of time-constancy---and so forbad an origin of the universe? \\
Obviously, if one is faced with trying to justify such symmetry principles, it is tempting to seek a general argument: for example, along the lines that: (i) the aims of science, or the formulation of  laws of nature, or some similar general goal, require or presuppose that ``Nature is uniform''; and (ii) this last requires spatial and-or temporal ``uniformity'', in a precise sense such as spatial isotropy, and spatial and-or temporal homogeneity.

So much by way of sketching the scientific, and philosophical, debates in cosmology some sixty years ago. To close this discussion, it suffices to note that these debates died away after the mid-1960s, owing to the refutation of the steady-state theory by the discovery in 1964 of the cosmic background radiation (CMB), the ``echo'' of the primordial fireball described by the Big Bang theory. Accordingly, the original objection above, that cosmology cannot be a science, also died away. Indeed, the fact that the CMB should still be detectable was deduced already in 1948 by  Big Bang theorists in their model of early-universe nuclear physics---but the prediction was forgotten about. Besides, this now-famous episode is itself a striking case of this Subsection's main point: that cosmology can manage perfectly well as a science, while invoking only the {\em laws} of local physical theories.\footnote{For a history of the 1950s debate about laws in cosmology, cf. Kragh (1996: Section 5.2, 219--51), who refers to philosophers such  as Dingle, Harre, Munitz and Whitrow, as well as physicists.  Massimi and Peacock (2015a) is a philosophical introduction. The prediction of the CMB was by Alpher and Herman (1948). For the history, cf. Kragh (1996: 132--5), Longair (2006: 319--23) and (more popular), Barrow (2011: 139--47), Singh (2004: 326--36, 428--37). Durrer (2015) is a technical review of (i) the history of investigating the CMB and (ii) the significance of the results, over the last fifty years. For a fine philosophical discussion of laws in  cosmology, cf. Smeenk (2013: Section 4). 

Note that nowadays `Big Bang' is ambiguous between three ideas: (i) the Big Bang theory, a very well confirmed theory of the evolution of the observable universe; (ii) the Big Bang fireball, i.e. the early conditions according to the Big Bang theory; and (iii) the Big Bang singularity, a hypothetical beginning of time which, as we will see in Section 5, is even more hypothetical in the context of eternal inflation. (Thanks to Anthony Aguirre for this point.)

Although we have no truck, and a scientific realist need have no truck, with Section \ref{laws?}'s original objection: its second, more scientific, question---how can we justify cosmological modelsÕ symmetry principles?---is undoubtedly important. For the CP, we will return to this briefly in Section  \ref{manchak}.}

\subsection{Ascertaining the global structure of spacetime?}\label{manchak}
We turn to reviewing some theorems (by Manchak (2009, 2011), building on ideas by Malament (1977) and Glymour (1977)) to the effect that, according to general relativity, one cannot know the global structure of space-time, even if one knew, as completely as one could,  the local facts about the structure of spacetime (and also, the local facts about the state of matter and radiation).

Our review is brief, for two reasons. (1): These theorems have already been discussed from a philosophical viewpoint (e.g. Beisbart (2009: 181), Norton (2011: Sections 5,6), Smeenk (2013: Section 5, 628--33) and Butterfield (2014: Section 2, 59--60). (2): These theorems---despite their foundational interest---have not influenced the  theoretical cosmology community. This is presumably because they are proved by a stupendous ``cut-and-paste'' construction on a given spacetime model: a construction which looks unphysical.\footnote{\label{bettworse}{But this lack of influence may well be unfortunate. As discussed in the references in (1), and in Manchak (2011): our defining a model by a cut-and-paste construction is no evidence at all that its features are not generic among general relativity's models.}}

These theorems provide a different perspective on the familiar theme of the under-determination of theory by data: the idea that all possible observations might fail to decide which of a set of  alternative theories is correct. For most philosophical discussions take ``all possible observations'' to mean the observations made by all observers, wherever situated in space and time, without regard to bringing the data together at some single point (or small spacetime region). But, on the other hand, cosmologists are continually confronted with the limits of our observational perspective on the universe: for example, that we can only now observe the past light-cone of Earth-now; and that direct observations by light, and other electromagnetic radiation, can go back only to the time (about 380,000 years after the Big Bang) when the universe first became transparent to radiation, i.e. to the last scattering surface, from which the CMB originates.  (On the other hand, we should note, indeed celebrate, examples where we break previous limits to our observational perspective: the obvious current example being the recent detection of gravitational waves (Abbott et al. 2016).)  Thus cosmologists will tend to be amenable to a definition of ``all possible observations'' which reflects such limitations---and these theorems work with just such a definition, albeit an idealized one.	

Thus Manchak envisages that an observer at a point $p$ in a spacetime $M$ might ascertain, by suitable observations, the metric structure of the past light-cone $I^-(p)$ of $p$: after all, information, such as measurement results, from within $I^-(p)$ can reach $p$ by a signal slower than light.  But Manchak goes on to prove that such an observer  cannot know much about the global structure of her spacetime, since many different spacetimes, with widely varying global 
properties, have a region isometric to $I^-(p)$ (where ``isometric'' means ``has the same metric structure as'').\footnote{It is usual to write $M$ for the spacetime, to indicate that it has the structure of  manifold; and the minus-sign superscript in $I^-(p)$ indicates the past, rather than future, light-cone. One might object that an observer could surely not ascertain so much as the metric structure of  her entire past light-cone. But in reply: (i) this idealization only makes the theorem stronger, along the lines ``even if you knew the metric structure of your entire past light-cone, you could not know the global structure''; (ii) there are theorems that support this idealization, e.g. about how to deduce the metric structure of the interior of the light-cone from information about its boundary (Ellis et al. 1985: Section 12).}

More precisely: let us take a spacetime to be a manifold $M$ equipped with a metric $g$, written $(M,g)$. Then Manchak defines a spacetime $(M,g)$ to be {\em observationally indistinguishable} from $(M',g')$ iff for all points $p \in M$, there is a point $p' \in M'$ such that $I^-(p)$ and $I^-(p')$ are isometric. (The fact that this
notion is asymmetric will not matter.) Then he proves that almost every spacetime is observationally indistinguishable from another, i.e. a non-isometric spacetime.

More precisely, the theorems  incorporate (i) a mild limitation; and  two significant generalizations ((ii) and (iii)).\\
 \indent (i): The theorems set aside spacetimes $(M,g)$ that are causally bizarre in the sense that there is a point $p \in M$ such that $I^-(p) = M$. (This last condition implies various causal pathologies, in particular that there are closed timelike curves.)\\
 \indent  (ii): But the theorems accommodate any further conditions you might wish to put on spacetimes, provided they are {\em local}, in the sense that any two spacetimes $(M,g)$ and $(M',g')$ that are locally isometric (i.e. any $p \in M$ is in a neighborhood $U \subset M$ that is isometric to a neighborhood $U' \subset M'$, and {\em vice versa}) either both satisfy the condition, or both violate it. (This means the theorems can probably be adapted to allow assumptions about the observer at $p$ ascertaining facts about matter and radiation in $I^-(p)$.)\\
 \indent  (iii): The theorems also prevent an observer's ascertaining some significant global properties of her spacetime. Manchak lists four such properties (2011: 413--414). (Three are ``good causal behavior'' properties: viz. that the spacetime be globally hyperbolic, inextendible and hole-free. We will not need their definitions. But it is worth noting the fourth property, spatial isotropy: there being, at every spacetime point, no preferred spatial direction. For this is  crucial to the cosmological principle mentioned in (1) at the end of Section \ref{laws?}.) Thus the theorems imply: given a spacetime $(M,g)$ with any or all of these properties, there is an observationally indistinguishable spacetime with none of them.

Thus Manchak's theorems (2009: Theorem, p. 55; 2011: Proposition 2) amount to the following. 
\begin{quote}
Let $(M,g)$ be a spacetime that is not causally bizarre,  that satisfies any set $\Gamma$ of local conditions, and that has any or all of the four listed global properties. Then there is a non-isometric spacetime $(M',g')$ such that: \\
\indent (a): $(M',g')$ satisfies $\Gamma$, but has none of the four listed global properties; and\\
\indent (b): $(M,g)$ is observationally indistinguishable from $(M',g')$.
\end{quote}
What should we make of this endemic under-determination of global spacetime structure, even  by perfect knowledge of the metric structure of the observer's past light-cone? 

Though we cannot discuss this at length, we stress that previous philosophical commentators (cf. the references at the start of this Subsection) are skeptical of the obvious realist strategy, viz. condemning some of the observationally indistinguishable alternatives as unphysical. In particular, there seems no good general reason to break the under-determination by imposing the cosmological principle (CP: cf. (1) at the end of Section \ref{laws?}). For many models that violate CP are physically reasonable (Beisbart and Jung 2006: 245--250; Beisbart 2009: Section 5, 189f.; Butterfield 2014: Section 3, 60--65). More generally, on the topic of justifying cosmological models' symmetry principles, we recommend: (i) for a conceptual introduction, Ellis (1975); (ii) for recent work on the prospects for showing the universe to be homogeneous, Clarkson and Maartens (2010) and Maartens (2011). 

However, as we also said: these theorems seem to have had no impact on  cosmologists. As we will discuss in Section \ref{know}, cosmologists take themselves to have established, during the last fifty years, a detailed account of the evolution of the universe, from less than a thousandth of a second after the Big Bang onwards. (Here, ``the universe'' can be understood as the past light-cone of Earth-now; or better: as the future light-cone of the past light-cone of Earth-now.) This is an account which endorses CP, by using a FRW model for spacetime. While many details of this account remain to be understood (for example:  the nature of dark matter and dark energy), there is a strong consensus about what has already been established: which we will describe in more detail in the next Section. Thus for cosmologists today, the live issues about under-determination relate---not to the global structure of the universe in the above sense (in particular, not to the rationale for CP), but---to:\\
\indent (i) ill-understood aspects of the account after the first thousandth of a second, e.g. our present evidence not settling the nature of dark matter and dark energy; and \\
\indent (ii) the history of the universe much (logarithmically!) before one second, especially the nature of the putative inflationary epoch, and the idea of a multiverse: i.e. issues about primordial cosmology: to which we will turn in Sections \ref{plan} and \ref{prob}. 

As mentioned at the end of Section \ref{bear}, these issues will threaten the usual distinction between under-determination by all data one could in principle obtain, vs. by all data obtainable in practice.

\section{A smidgeon of what we now know}\label{know}
The last fifty years have seen the triumphant rise of observational cosmology as a precision science. Looking back, the discovery in 1964 of the cosmic background radiation (CMB) was an iconic initial event. But at least as important was the subsequent use of satellites  to make precise observations about parts of the electromagnetic spectrum that we cannot observe on earth: i.e. neither optical nor radio wavelengths. This of course includes the microwave wavelengths of the CMB itself: whose properties have been measured with ever-greater precision, probing ever more finely the early universe, by a  succession of satellites: COBE, WMAP and  Planck. This  has also of course depended on inventing and developing sensitive yet robust instruments that can be reliably operated remotely, i.e. by radio signals: an extraordinary collaborative achievement of diverse disciplines, ranging from quantum optics to software engineering. And quite apart from the CMB, the last fifty years has witnessed various large and detailed observational programs, both terrestrial and by satellite. The eventual upshot of these observations,   building up from countless precise measurements  of many diverse quantities---relating not just to the cosmos as a whole but to the structure of stars and of galaxies---has been to give us a detailed overall history, not just of the evolution of the universe as a whole, but also of the formation  of galaxies and the life-cycles of stars. Undoubtedly, we live in a golden age of cosmology! 

This is not the place for a detailed review of these developments aimed at philosophers, fascinating though they are: both as regards the scientific methods used and the stupendous cosmic history thus inferred.  We have already, in (1) of Section \ref{bear},  ducked out of the attempt to say exactly which results of modern cosmology---which chapters of the overall story just mentioned---should count as ``mature and accepted'' and so, for we scientific realists, as approximately true in a correspondence sense.\footnote{So we here must set aside several  topics that, independent of the general effort to assess how scientific realism fares in cosmology, would form excellent case-studies. For example: (i) the life-cycles of stars, for which cf. e.g. Chandrasekhar (1939), Kaler (2006), Ryan and Norton (2010);  (ii) the sophisticated methods and instruments used e.g. to establish astronomical and cosmological distances, for which cf. e.g. Pasaschoff (1994), Rowan-Robinson (2011: Chapter 3) and Longair (2003: Chapter 18, 478--98; 2006, Chapters 7, 11, 13).} 

But to set the scene for later Sections' discussions, we need to cite a few `bare bones' of the overall thermal history of the universe (Section \ref{four}). This will lead in to a brief defense of scientific realism in relation to cosmology (Section \ref{sr}).  

\subsection{Four snapshots of the early universe \dots }\label{four}
We have already mentioned that the properties of the CMB confirmed the idea of a primordial fireball described by the Big Bang theory. More precisely, the idea is that all the matter in the entire universe we can observe today was once, about 13.8 billion years ago,  confined to a much smaller volume, at energies, temperatures and densities so high that atomic structure breaks down, and there is instead a ``soup'' of subatomic particles. This idea has been worked out in detail and with great quantitative precision, by combining various parts of established physics: the physics of how subatomic particles interact (comprising both the standard model of elementary particle physics, and nuclear physics); the general relativistic description of spacetime; and (relativistic) thermodynamics and hydrodynamics. 

Agreed: as one considers earlier times, the energies (and temperatures and densities) get so high that discussion needs must go beyond established physics. But amazingly, established physics suffices to describe the fireball right back to about $10^{-4}$ seconds after the Big Bang!\footnote{{\label{-4}}{We here take ``{\em about} $10^{-4}$ seconds'' as the ``cut-off for credence'', not because of a specific problem or controversy, but simply because reason and evidence do not dictate a unique cut-off: recall the discussion in (1) of Section \ref{bear}. Certainly, $10^{-4}$ seconds is endorsed by some authorities; e.g. Rowan-Robinson (2011: 100), and forty years ago, Weinberg endorsed $10^{-2}$ seconds (1977: 5). But as we discuss below: it is also reasonable to be less cautious, since it is common nowadays to take the boundary between known and speculative physics to be at about  $10^{-11}$ seconds (!) after the Big Bang as the cut-off (e.g. Earman and Mosterin 1999: 2). Agreed: it is also reasonable to be more cautious, even taking one second.  Anyway, for the interests and purposes of philosophers, the scientific story is equally amazing when one is more cautious: say, withholding credence for times earlier than one second. And certainly it is not reckless for a scientific realist to endorse the story from about this time: one renowned expert says he is 99\% confident of the story from one second onwards  (Rees: 1997: 65, 17; 2003: 24, 31).}}

We now spell this out a bit. Two preliminary points: in effect, the first is about space, and the second about time.\\
\indent (1): We referred to ``the entire universe we can observe today''. Bearing in mind that light has a finite speed, so that we see more distant objects as they were at earlier times, the appropriate meaning of this phrase, and similar phrases like ``the observable universe'' is: the past light-cone of Earth-now. So the idea of a primordial fireball is that all the matter and radiation content of the past light-cone of Earth-now was once confined to a much smaller volume.\footnote{Notice that on this usage, ``observable'' in ``the observable universe'' does not connote being macroscopic: the observable universe includes all physical objects and events in the past light-cone of Earth-now, no matter how microscopic. This usage was of course also in play in Section \ref{manchak}'s idealization that an observer at $p$ could ``observe'' metric structure on arbitrarily fine length-scales, albeit only within her past light-cone.}\\
\indent (2):  It will be helpful (though difficult!) to think logarithmically, not arithmetically: to think, for example, that since the present time is about $10^{17}$ seconds after the Big Bang, the time  $t = 10^{-17}$ seconds before the Big Bang is {\em as much before $t$ = 1 second, as we are after it}. Though this sounds blatantly wrong, the rationale for it is that, as every physicist knows, physics is a matter of scales, i.e. orders of magnitude. That is: if you change the object, or topic, or regime, you wish to describe by an order of magnitude, i.e. by a {\em factor} of about 10, you are liable to need a very different description: and even more likely if you change by two orders of magnitude, i.e. by a factor of about 100. This trend holds whether the quantity whose value you change is time, or distance, or energy or temperature: (indeed in quantum theory and relativity theory, these quantities are intimately related, so that changing one involves changing others). In particular, this means that when cosmologists puzzle over what was the state of (the matter and radiation now comprising) the observable universe, at say $t = 10^{-6}$ seconds, or how physical processes changed as a result of the cooling between, say,  $t = 10^{-11}$ and $t = 10^{-6}$ seconds, we should not accuse them of straining at gnats, i.e. of myopically concentrating on very transient matters which cannot matter very much. For, agreed: the universe was changing unbelievably rapidly (arithmetically speaking!); but the relevant processes change---and so our description must change---in crucial ways, depending logarithmically on the earlier time considered.\\

So here are some snapshots from the thermal history of the observable universe: four snapshots, in reverse chronological order, all corresponding to established physics.\footnote{{\label{thermal}}{Agreed, only the first two concern times later than $t = 10^{-4}$ seconds: which in footnote \ref{-4}, we took as the cut-off for credence. But recall that this choice was in the middle of the reasonable spectrum, from $10^{-11}$ seconds to one second: cf. also the ensuing discussion in the main text. 

Another indicator of how well-established is this thermal history is the fact that authoritative textbook descriptions of it, written over the last forty years, largely agree with each other. Cf. for example: Sciama (1971: Chapters 8, 12--14), Weinberg (1972: Chapter 15.6, pp. 528--545), Wald (1984: 107--117), Barrow and Tipler 
(1988: 367--408, Sections 6.1--6.7), Lawrie (1990: 315--326), Longair (2006: 394--399), 
Weinberg (2008: 101--113, 149--173; Sections 2.1, 2.2, 3.1, 3.2). For fine popular 
accounts, cf. Weinberg (1977, Chapters 5,7), Silk (1989: Chapters 6 to 8), Rowan-Robinson (1999: Chapter 5),  Silk (2006: 112--128).}} 
 
 \begin{quote}
(A). $t = 10^{13}$ seconds: which is about 380,000 years after Big Bang. This is an important time: for at about this time, the universe first became transparent to radiation (by free electrons combining with nuclei to form atoms). So our direct observations by light, and other electromagnetic radiation, can go back only to this  {\em decoupling time} $t_{dec}$. (It is also known as the time of {\em recombination}: though all agree that  {\em combination} would be a much better name, since the electrons and nuclei were not stably combined at any earlier time.) The temperature is about $3000$ K (corresponding to an energy of  $\sim 0.3$ eV). (By way of comparison, the temperature at the surface of the Sun is about $6000$ K). The size of the universe relative to its size today (as given by the appropriate ratio of the scale factor) is $\sim 10^{-3}$, and the mass density is $\sim 10^{-21}$ $\textrm{g}/\textrm{cm}^{3}$.\\
\\
(B). $t = 10^{-2}$ seconds after the Big Bang. Temperature $\sim 10^{11}$ K (corresponding to an energy of  $\sim 10$ MeV). Nuclei form: i.e. at higher temperatures, they ``melt'' into their constituent protons and neutrons. The size of the universe relative to its size today is $\sim 10^{-11}$, and the mass density is $\sim 10^{9}$ $\textrm{g}/\textrm{cm}^{3}$.\\
\\
(C). $t = 10^{-6}$ seconds after the Big Bang. Temperature $\sim 10^{13}$ K (corresponding to an energy of  $\sim 1$ GeV). Protons and neutrons form: i.e. at higher temperatures, they ``melt'' into their constituent quarks.The size of the universe relative to its size today is $\sim 10^{-12}$, and the mass density is $\sim 10^{17}$ $\textrm{g}/\textrm{cm}^{3}$.\\
\\
(D). $t = 10^{-11}$ seconds after the Big Bang. Temperature $\sim 10^{15}$ K (corresponding to an energy of  $\sim 100$ GeV). This is the temperature/energy above which the electromagnetic and weak forces are in a sense unified in the electro-weak force (viz. by the effective potential being SU(2)-symmetric), and below which they are distinguished by the Higgs mechanism. That is: this is the temperature/energy at which the Higgs mechanism works, producing an electro-weak phase transition. Since the discovery of the Higgs particle in 2012, this can be taken as the upper end of the confirmed energy range for the  standard model of particle physics. That is:  before $t = 10^{-11}$ seconds, the energies are too high for us to be confident that the  standard model  applies. The size of the universe relative to its size today is $\sim 10^{-15}$, and the mass density is $\sim 10^{27}$ $\textrm{g}/\textrm{cm}^{3}$.
 \end{quote}
We emphasize that this thermal history is by no means the only main claim of modern cosmology that is now firmly established. Other examples include our theories of stars and galaxies: for which one could, as in footnote \ref{thermal}, cite authoritative descriptions over several decades which largely agree with each other. But more relevant to us is the way in which the CMB, though very homogeneous and isotropic, has tiny irregularities whose structure (especially how their size depends on angle) gives detailed evidence about the interplay {\em before} the decoupling time, between gravitation tending to clump the matter, and radiation pressure opposing the clumping. This interplay links directly to parameters describing the universe as a whole, such as whether the average density is large enough for gravitational attraction to eventually overcome the expansion, so that the universe ends in a Big Crunch instead of expanding forever. In the last fifteen years, data about these irregularities gathered mostly by satellite projects like WMAP and Planck have been used to estimate these cosmological parameters, resulting in a striking concordance with estimates obtained by completely different methods. We will return to this topic in Section \ref{plan}.

So much by way of citing some of modern cosmology's stupendous claims about the very early universe. Let us return to our warrant for believing them, and thus our defense of scientific realism. 

\subsection{\dots as viewed by a scientific realist}\label{sr}
Our main view is (as we said in (1) of Section \ref{bear}) that indeed, cosmology has definitively established such stupendous claims as those cited at the end of Section \ref{four}. But we should add three clarifications to this realist {\em credo}. They expand on our basic point in Section \ref{intro}, that while scientific realism claims, roughly speaking, ``we can know, indeed do know, about the unobservable'', it does {\em not} claim that ``all the unobservable is known''; or even, ``all the unobservable is knowable''.\footnote{For a complementary discussion, cf. Butterfield (2012: Section 2.2).}\\

(1): {\em A spectrum of reasonable credence}:--- Scientific realism enjoins us to believe propositions about the unobservable only when the evidence is sufficiently plentiful and varied. Of course, there can be no general statement of what would be ``sufficient''. And for cosmology, all parties admit that the evidence gets thinner as we consider earlier and earlier times, corresponding to ever-higher energies, temperatures and densities. As we mentioned in Section \ref{four} (footnote \ref{-4} and snapshot (D)), it is common  nowadays to take the boundary between known and speculative physics to be  at about $10^{-11}$ seconds after the Big Bang (this time corresponding to the electro-weak phase transition). So for earlier times, observational data are so lacking and theory is accordingly so speculative, that one has to be agnostic. 

But of course, there is a spectrum here: of credence, as well as times and energies. The standard model has indeed passed every test that experimental high energy physicists have subjected it to, in the forty years since its formulation in the mid-1970s; (culminating in the discovery of the Higgs boson in 2012). And the theories of nuclear physics, that describe the synthesis of nuclei from protons and neutrons (a lower-energy process occurring at about $10^{-2}$ seconds after the Big Bang) are even better confirmed than the standard model. But of course not every aspect of the standard model (or even of nuclear physical theories) has been confirmed: especially, its description of phenomena at the upper end of its (impressively wide) energy range. So a scientific realist might be cautious, and not believe the standard model's description of the very early universe, for times earlier than some cut-off time that is {\em later} than $10^{-11}$ seconds. For example, they might have a cut-off time as late as $10^{-2}$ seconds, or one second: thus being more cautious than we reported in footnote \ref{-4}.

On the other hand, a less cautious attitude is also reasonable. Aspects of the standard model involved in `snapshots' (C) and (D) above, i.e. for $t = 10^{-6}$ and $t = 10^{-11}$ seconds, {\em have} been confirmed in recent {\em terrestrial} experiments.\\
\indent (i):  Corresponding to snapshot (C): in the RHIC (Relativistic Heavy Ion Collider) at Brookhaven, USA, protons in gold nuclei have been ``melted'' to produce a (very short-lived!) quark-gluon plasma; (for an experimental review, cf.  e.g. Shuryak 2005). This means the quarks were liberated from their confinement inside a proton, after some $13 \times 10^9$ years: indeed, a long prison sentence! \\
\indent (ii): Corresponding to snapshot (D): at the LHC (Large Hadron Collider) at CERN in 2012, the discovery of the Higgs particle lent credence to the mechanism, proposed in the standard model, whereby interactions with the Higgs field gives elementary particles their mass. Thus it would  also be reasonable to take the standard model, even at the energies obtaining at $t = 10^{-11}$ seconds, to be confirmed by terrestrial experiments: and to be a ``mature and accepted'' theory that earns our belief.  So a scientific realist could  take $t = 10^{-11}$ seconds as the cut-off for credence.

To sum up: in the last forty years, we have been very fortunate, as regards both: (a) confirming the standard model in particle accelerators; and (b) successfully applying it to the early universe---whose enormous density makes it a very different regime from the near-perfect vacuum of an accelerator.\footnote{Our success in (b) depends on a striking feature of the theory of quarks and gluons, viz. asymptotic freedom: roughly, the strength of their interactions decreases at high density.}  \\

(2):  {\em Towards terra incognita}:--- But it is also reasonable to fear that this good fortune cannot continue! More precisely: we should expect that, however fortunate we {\em may} be in the future, as regards both (a) making observations and (b) developing theories, we shall never know all, or even much, about {\em arbitrarily} high energies and thus about arbitrarily early times. There are really three points in play here: two about high-energy physics, and one about cosmology.\\
\indent (i): As regards making observations, we cannot expect to---we cannot afford to!---build particle accelerators that achieve arbitrarily high energies: or even energies much above those now attainable. (But we should also note, in  a more optimistic tone, the power of human ingenuity and tenacity to make quantitative observations of the most arcane kind.  The obvious current example is again the recent detection, culminating forty years of effort, of gravitational waves (Abbott et al. 2016): that is, the detection of distortions of spacetime on a length-scale of a thousandth of the diameter of a proton!)  \\
\indent (ii): As regards developing theories, we now realize (largely as a result of our modern understanding of renormalization, led by Wilson's work from the mid-1960s) that by and large, the most we can hope for is {\em effective} theories: i.e. theories that accurately describe phenomena that occur in a certain energy range, but are inaccurate for higher energies. \\
\indent (iii): As regards cosmology, recall that ``the Big Bang'' is really a label for {\em terra incognita}. Formally, it labels a singularity of our theoretical descriptions, both the general relativistic description of spacetime (infinite curvature etc.) and the quantum description of matter and radiation (infinite energy, density etc.). But these infinities surely represent breakdowns of our theories, not physical realities. And as we consider higher and higher scales, of curvature, or of energy or density etc., we have no guarantee that we are capable of formulating an accurate theory for phenomena at those scales---or even that our basic concepts such as spacetime, and physical quantities like energy and momentum, apply at those scales. And bearing in mind that we need to think logarithmically: it is no solace to be told that this cognitive lacuna is over in a minuscule fraction of a second!\footnote{For some more details, for philosophers, about point (ii), cf. e.g. Butterfield (2014a), Butterfield and Bouatta (2015). Point (iii) broaches the vast topic of our search for a quantum theory of gravity: for introductions aimed at philosophers, cf. e.g. Butterfield and Isham (2001), Rovelli (2007), Rickles (2008), and Dawid, this volume.

Point (iii) also broaches two more specific topics: (a) the need to scrutinize whether our concepts of space and time apply; cf. issue (2) in the list at the end of Section \ref{intro}, and the work of Rugh and Zinkernagel: (b) the need to scrutinize the definition of, and the occurrence of!, singularities in general relativity. For (b) we recommend Curiel (1999). We also thank him for emphasizing that one should not blithely claim that quantum effects will efface singularities, or that singularities will not appear in an ultimate theory of quantum gravity. For notice that Wall (2013) shows that if the Generalized Second Law is valid, then there will necessarily be singularities even in regimes where quantum effects make themselves felt.}\\

(3): {\em Conceptual change}:--- Our realist {\em credo} is not intended to deny major epistemic ruptures, such as are often dubbed `conceptual change' and-or `meaning variance'. Progress in cosmology has of course involved such ruptures, in the process of establishing the present consensus about the universe's history after about a thousandth of a second: the outstanding example from twentieth-century cosmology is, no doubt, general relativity's description of spacetime as dynamical, and of the universe beginning in a singularity. And future research about that history will presumably  again involve such ruptures. Looking ahead, the obvious putative examples are dark matter and dark energy. They are a major causal and structural aspect of the cosmic history revealed by modern cosmology: indeed, they  dominate the mass-energy content of the universe. But their nature is not understood: and gaining that understanding may involve some kind of epistemic rupture. So agreed: we still have a lot to learn. 

But a great deal {\em is} now established. In particular: we can be confident (albeit not certain!) about the thermal history of universe, as sketched in the four snapshots above. For we have very good physical reasons to believe that  this history is robust (especially at times from about  a thousandth of a second onwards) to whatever the dark matter and dark energy turn out to be.

\section{The very early universe: inflation?}\label{plan}

In the remainder of this paper, we turn to questions about the {\em very} early universe:  roughly speaking, about  times earlier than  $t = 10^{-11}$ seconds, which as we saw in Section \ref{know} represents the ``boundary'' of our confirmation of the standard model of particle physics. We will confine ourselves to the most widely accepted framework for understanding this regime: inflation.\footnote{\label{InflSceptm}{But we stress at the outset that inflation remains a speculation, and that there are various respectable alternatives. For {\em maestri} being skeptical about inflation, we recommend:  Ellis (1999, p. 706--707; 1999a, pp. A59, A64--65; 2007, Section 5, pp. 1232--1234), Earman (1995, pp. 149--159), Earman and Mosterin (1999), Hollands and Wald (2002), Penrose (2004, pp. 735--757, Chapters 28.1--28.5), Steinhardt (2011), and Ijjas et al.~(2013); though we of course also recommend replies, such as Guth et al. (2014). For reviews of some alternatives, such as string gas cosmology, cf. e.g. Brandenberger (2013), or, aimed at philosophers, Brandenberger (2014).}}  Thus Section \ref{idea} begins by introducing inflation, and so functions as a prospectus for both this Section and Section \ref{prob}. Section \ref{cause} discusses the conjectured mechanisms for it, and  Section \ref{pleth} presents this Section's main point about scientific realism: that the details of the mechanism are seriously under-determined by the data.

\subsection{The idea of an inflationary epoch}\label{idea}
As mentioned in Section \ref{intro}: from about 1980 onwards, cosmologists have proposed that there was a very early (and so brief!) epoch of very rapid, indeed accelerating, expansion. They have made three main claims about this epoch: claims which will dominate this Section and the next, so that we give them mnemonic labels.\\
\indent \indent (Three): Such an epoch solves three problems faced by the existing Big Bang model, which by 1980 was well-established (it had already been dubbed ``standard'' by Weinberg (1972: 469) and Misner et al. (1973: 763)). These problems are the ``flatness'', ``horizon'' and ``monopole'' problems.\\
\indent \indent (Inflaton): If this epoch was appropriately caused---viz. by a conjectured inflaton field---it would lead to characteristic features of the CMB: namely, characteristic probabilities for the amplitudes and frequencies of the slight wrinkles (unevennesses) in the CMB's temperature distribution. \\
\indent \indent (Branch): This mechanism for inflation would naturally involve a branching structure in which, during the epoch, countless spacetime regions branch off and themselves expand to yield other universes; so that the whole structure is a ``multiverse'', whose component universes cannot now directly observe/interact-with each other, since they are causally connected only through their common origin in the very early universe.  

The evidence for, and status of, these three claims varies. Most cosmologists regard claim (Three) as established: i.e. there was an epoch of accelerating expansion (however it may have been caused) and the occurrence of this epoch  solves the flatness, horizon and monopole problems. But cosmologists agree that the cause of this epoch---its mechanism: the dynamical factors that started it, and then played out so as to end it---is much more conjectural; and therefore, so also is the third claim, (Branch).\footnote{\label{WeinbWarn}{Thus Weinberg says: ``So far, the details of inflation are unknown, and the whole idea of inflation remains a speculation, though one that is increasingly plausible'' (2008: 202).}} The original proposal was that the mechanism was a scalar field, $\phi$, the inflaton field, evolving subject to a certain potential $V(\phi)$; and that this mechanism led to characteristic features of the CMB. It is testimony to the strength of this proposal that it remains the most popular mechanism, and there remain versions of it which are confirmed by the CMB data. But this mechanism, and so the claim (Inflaton), is undoubtedly more speculative than the mere occurrence of the epoch of expansion. So we must beware of the ambiguity of the word ``inflation'': it can refer either to the epoch of expansion (also called ``inflationary epoch''), however it was caused; or to the (more speculative) mechanism for its occurrence. Finally, the idea that the mechanism for inflation spawns many universes, i.e. claim (Branch),  is even more speculative.
 
We will now concentrate on the least controversial claim, (Three): i.e. the claim that there was an inflationary epoch (however it was caused) and that such an epoch solves the three problems---flatness, horizon and monopole---faced by previous general relativistic cosmological models. Then Sections \ref{cause} and \ref{pleth} take up claim (Inflaton), about what caused the inflationary epoch. And Section \ref{prob} will take up claim (Branch), about the multiverse. 

\subsubsection{An inflationary epoch solves three problems}\label{3}
The way in which an inflationary epoch solves the three problems has already attracted philosophical discussion (e.g. Earman (1995: Chapter  5, especially 142--159), Earman and Mosterin (1999: Sections 4--7, 14--26), Smeenk (2013: Section 6, 633--638), Butterfield (2014: Section 4.1--4.2, 65--67), McCoy (2015)). This is not least because two of the three problems are problems, not so much of the empirical adequacy of general relativistic cosmology, as of {\em explanation}---a natural topic for philosophers. 

Thus, cosmology's established model in the mid-1970s has two features, each of which looks like an implausible coincidence that cries out for explanation. In both respects, the model is empirically adequate: the feature in question does not contradict observations. But the feature requires an aspect of the model that is otherwise free, i.e. not constrained by theory, to be fine-tuned to an extreme degree, on pain of empirical inadequacy. The simplest and clearest example is the {\em flatness problem}, where the aspect concerned is a theoretically central parameter about the spacetime geometry; and it has to be fine-tuned to many decimal places. So we shall give some details about that; then we will mention how similar considerations apply to the {\em horizon} and {\em monopole problems}.

 According to the Big Bang models of the 1970s (based on the Friedmann-Robertson-Walker (FRW) metric), there are three main possibilities for the fate of the universe.  Either:\\
 \indent \indent (i): the universe's matter and radiation is on average dense enough that gravitation will eventually overcome the expansion, so that the universe is fated to reach a point of maximum size and then ``turn around'', i.e. contract and end in a Big Crunch (and of course: the greater the density, the sooner will the turn-around and Crunch occur): (called a {\em closed universe}); or \\
 \indent \indent (ii): the matter and radiation density is low enough that gravitation cannot overcome the expansion, though it slows the rate of expansion down to some asymptotically non-zero value: (called a {\em open universe}); or \\
 \indent \indent (iii): the matter and radiation density is (a) low enough that gravitation cannot overcome the expansion, but (b) high enough that it slows the rate of expansion down, asymptotically,  to {\em zero}: (called a {\em flat universe}, since the spatial  geometry  of the instantaneous spatial slices is asymptotically Euclidean).
 
Obviously, (iii) i.e. flatness is  theoretically privileged. For this value of the density represents the boundary between the regimes (i) and (ii).  So the important dimensionless number is the ratio---which is written as $\Omega$---of the actual density to this critical value. So $\Omega = 1$ corresponds to flatness.  

According to the FRW models, if $\Omega$ is ever 1, then it is always 1. And in fact, the observed value of $\Omega$ is close to 1. The universe is now, and indeed at all times later than about one second after the Big Bang, almost flat: its spatial geometry is almost Euclidean. But in these models, any difference of $\Omega$ from 1 in the early universe is very rapidly amplified.   For example: if at one second after the Big Bang, $\Omega = 1.08$, then already at ten seconds $\Omega = 2$; and thereafter $\Omega$ keeps increasing exponentially. And on the other hand: if at one second after the Big Bang, $\Omega = 0.92$, then already at ten seconds $\Omega = 0.5$; and thereafter $\Omega$ keeps decreasing exponentially. In short: $\Omega = 1$ represents an equilibrium---but a very unstable equilibrium. And this means that for $\Omega$ to be about 1 today requires that it be stunningly close to this privileged value  soon after the Big Bang: for example, it has to be $1 \pm 10^{-16}$ at one second after the Big Bang.\footnote{For these numbers, cf. Guth (1997: 25), or Liddle (2003: 100). Note that $ 10^{-16}$ is about the ratio between the width of a human hair (viz. a tenth of a millimeter) and the average distance between Earth and Mars (viz. 225 million kilometers)! And of course, for times earlier than one second, the fine-tuning is to yet more decimal places.} 

Should this degree of fine-tuning be treated as a ``coincidence'': i.e. a brute fact, set aside from the quest for explanation? Agreed: it is in general a matter of judgment, not dictated by the scientific context, what facts it is legitimate to treat as brute. But in this case: the fine-tuning is so extreme that it cannot be dismissed as brute.

And the idea of an inflationary epoch promises to predict and-or explain $\Omega$ being close to 1. For it is easy to show that whatever the value of $\Omega$ at the onset of the inflationary epoch, $\Omega$ will be driven close to 1 by the end of the period, and will remain close to 1 for a {\em very} long time thereafter---including until now. (So the explanation is dynamical: in philosophers' jargon, causal.) Besides, the idea of this explanation is very simple: an expansion of a highly curved surface makes a local patch flatter. Think of blowing up a balloon; or how the fact that the earth is large makes our local patch of it seem flat (e.g. Guth 1997: 176--177;  Liddle 2003: 99, 104; Serjeant 2010: 27--28, 56). 
 
The horizon problem has a similar structure. Namely: our cosmological description (as of the late 1970s) is empirically adequate, but requires an aspect---in this case, an initial condition---to be fine-tuned to an extreme degree: so extreme that it is implausible to treat it as a brute fact. Recall that the CMB, dating from the decoupling (recombination) time $t_{dec}$ (380,000 years, or about $10^{13}$ seconds), is very homogeneous and isotropic. Its wrinkles are minuscule: their proportional size is $10^{-5}$, which is like having, on the surface of a pool of water one meter deep,  a wave only a hundredth of a millimeter high! Yet for two directions in space with a sufficient angular separation, the two past events at the time $t_{dec}$ that lie along those directions , i.e. the two emission-events of the CMB, have---according to the 1970s' models---no common causal past. That is, their past light-cones do not intersect. In the notation of Section \ref{manchak}, calling the events $p, q \in M$: $I^-(p) \cap I^-(q) = \emptyset$. This means there can be no process of thermalization or equilibration---or any kind of interaction---establishing the strong correlation in the properties of the CMB coming from the different directions. The strong correlation must just be accepted as a brute fact. This is all the more embarrassing when one calculates that the angular separation sufficient to imply no common causal past is tiny: about 2 degrees.\footnote{Embarrassment at having to accept strong correlation as a brute fact echoes the intuitive plausibility of Reichenbach's Principle of the Common Cause: which is much discussed as a motivation for the locality assumptions in Bell's theorem (cf. Ruetsche, this volume). For more about this comparison, cf. e.g. Earman (1985: Chapter 5), Butterfield (2014: Section 4.2).} 

Again, the idea of an inflationary epoch solves the problem: and in a simple way. Namely: a suitable inflationary epoch implies that the past light-cones of all emission events of the CMB---even for points on opposite sides of the sky---do in fact intersect; and so there could have been a suitable process of equilibration. In this way, the uniformity of the CMB, even on opposite sides of the sky, is explained: (e.g. Guth 1997: 182--186;  Liddle 2003: 102, 109; Serjeant 2010: 53, 55). 

Turning to the monopole problem:  this is a problem of empirical adequacy, rather than a problem of having to treat as brute what seems to need an explanation. It also differs from the flatness and horizon problems in being a matter of detailed, albeit speculative, physics, rather than spacetime geometry. Namely: there is good reason to think that at sufficiently high energies (and so: sufficiently early times) the strong and electro-weak forces are unified, in the same sort of way that classical electromagnetism {\em \`{a} la} Maxwell unifies electric and magnetic forces; (recall snapshot (D) in Section 3.1). Theories of this unification, which began to be formulated in the 1970s, are called ``GUTs'' (for ``Grand Unified Theories''). The problem is that GUTs, applied to the very early universe, predict  the production of very many magnetic monopoles: these are particles that have a magnetic charge, broadly analogous to the familiar electric charge. But these particles have never been detected: a matter of contradiction with observation (rather than embarrassment at treating a fact as brute). 

As to how an inflationary epoch solves this problem:  the idea of the solution is attractively simple, as it was for the flatness and horizon problems. Namely: (i) one accepts that the magnetic monopoles are produced, thus not questioning the admittedly speculative physics; but (ii) the enormous expansion occurs after, or at worst during, monopole-production, so that the expansion vastly {\em dilutes} the monopoles. Thus in the tiny pre-inflation patch that expanded to become the observable universe, there would be so few magnetic monopoles that we would not expect to have detected them.\footnote{Our description of  the monopole problem is so brief that it may be misleading: but a detailed description inevitably leads into the issue of what caused the inflationary epoch, i.e. claim (Inflaton) which we have postponed to Section \ref{cause}. For a superb popular account, by one of the inventors of inflation, cf. Guth (1997: Chapter 9, 147--165). A detailed account of inflation's solutions to all three problems can be found in Kolb and Turner (1990: Section 8.1--8.2) and Linde (1990: Section 1.5--1.7).} \\

So much by way of presenting the three problems; and the simple underlying ideas of the solutions given by an inflationary epoch.  This discussion prompts the question: {\em how much inflation?} That is: how much expansion of the universe, involving how much acceleration---and when---makes the three solutions come out {\em quantitatively} correct? 

Of course, each problem prompts a separate calculation of how much expansion, and when, would be sufficient for the solution. Needless to say, each calculation involves auxiliary assumptions, which can be questioned. But the calculations are also partly independent of each other: so the fact that the resulting estimates agree pretty well is  evidence in favor of an inflationary epoch.  We of course cannot give details of such calculations. We just report the consensus that all three solutions come out quantitatively correct, if we postulate figures like the following (dizzying though they be!):\\
\indent \indent (a): the inflationary epoch ends at about $10^{-34}$ seconds: which is a time corresponding to GUT-scale energies, viz. $E \approx 10^{15}$ GeV (so that the temperature $T = E/k_{\scriptsize\textrm{B}} \approx 10^{28}$ K);\\
\indent \indent (b): the inflationary expansion is exponential: and started at, for example,  $5 \times 10^{-35}$ seconds with a characteristic expansion time (i.e. the time in which the scale factor---the radius of the universe---is multiplied by $e \approx 2.7$) of about $10^{-36}$ seconds.\\
Taken together with (a), this would imply expansion by a factor  $e^{50} \approx 10^{22}$! (For further details, cf. e.g. Liddle 2003: 106--7; Serjeant 2010: 55.)

Let us sum up this Subsection's discussion of claim (Three). It is generally agreed that postulating an inflationary epoch, occurring in a suitable time-interval and increasing the size of the universe by suitably many orders of magnitude, solves the flatness, horizon and monopole problems. All three solutions invoke a satisfyingly simple idea so as to explain the puzzling feature in question. Besides, these solutions are resilient, in that they are independent of what might have caused the inflationary epoch ... but of course, one still asks: what caused that epoch?

\subsection{What caused the accelerating expansion?}\label{cause}

In answering this question, the first thing to stress is that we are here in the realms of  speculation: recall Weinberg's warning in footnote \ref{WeinbWarn}! 
We shall confine ourselves to reporting a ``minimal'' answer: which postulates  a scalar field, $\phi$, the inflaton field, evolving subject to a certain potential $V(\phi)$---cf. claim (Inflaton) at the start of Section \ref{idea}. As we will see, a few assumptions about this mechanism yields some characteristic predictions for subtle features of the CMB. And  since these features have  been observed by a sequence of increasingly refined instruments (such as the satellites COBE, WMAP and Planck mentioned at the start of Section \ref{know}), these confirmed predictions are nowadays regarded as a more important confirmation of inflation, than its solution of Section \ref{3}'s three problems. There is, however, a very considerable under-determination of the mechanism of inflation by our data---both today's data, and perhaps, all the data we will ever have. This (unfortunate!) predicament will be the topic of Section \ref{pleth}.  (Then in Section \ref{prob} we will turn to the most controversial claim (Branch), that inflation yields a multiverse of domains.)

\subsubsection{The inflaton field}\label{field}

We begin with a simple classical picture of inflation, postponing quantum considerations to Section \ref{SEC:CMB}. We begin by assuming that at approximately $10^{-35}$ seconds after the Big Bang, the stress-energy of the universe was dominated by that associated with some (yet to be discovered) scalar field $\phi(t,\vec{x})$, known as the \emph{inflaton}; and that the evolution of this scalar field is determined by a potential energy density $V(\phi)$, and by its coupling to the gravitational field.

We also assume that the dynamics of the scalar field is very simple, as follows. The scalar field is homogeneous: i.e., the same throughout space, so that the field is then only a function of time, $\phi(t,\vec{x})\equiv\phi_{0}(t)$.\footnote{Agreed, you might object that assuming homogeneity undercuts the claim to have solved the horizon etc. problems. But in reply: there is active research to ascertain how much homogeneity is needed to secure inflation (Brandenberger (2016), East et al. (2015), Kleban and Senatore (2016)); and anyway, we aim in this essay only to lay out the main ideas.} And the potential has a single minimum towards which the field `rolls': much as a classical particle would roll towards the minimum of a potential, subject to a force in the direction of the negative of the potential's gradient. 

It turns out that accelerating expansion of the underlying spacetime can occur when the potential energy density dominates over the kinetic energy of the inflaton field: i.e. $V(\phi_{0})  \gg \frac{1}{2}\dot{\phi}^2_{0}(t) $: a regime called the \emph{slow-roll} of the inflaton.  In a bit more detail: what one requires for accelerating expansion is a fluid with a pressure $P$ that is negative; more specifically, $P<-\frac{1}{3}\rho$, where $\rho$ is the energy density of the fluid. For the scalar field above (which can be thought of as a fluid in this sense), where $V(\phi_{0})  \gg \frac{1}{2}\dot{\phi}^2_{0}(t)$, the pressure $P$ comfortably satisfies this constraint. 

Inflation comes to an end when the inflaton finds its way to the minimum of the potential; typically, the slope of the potential gets larger, and the kinetic energy of the inflaton increases so that accelerating expansion can no longer occur. The inflaton oscillates (with a decaying amplitude) around the minimum of the potential, and the stored energy is released into particles of the standard model of particle physics through a process known as \emph{reheating}. 

Although this is just one means through which the universe can expand by the amounts needed to solve the three problems discussed in Section \ref{3}, taking this slow-roll  scenario seriously---that is, analyzing single-field slow-roll inflationary models, and deducing observational parameters---has hitherto been cosmologists' predominant way of exploring and assessing the idea of inflation.

\subsubsection{Connecting the inflaton to the CMB}\label{SEC:CMB}

If this was all there was to the theory of inflation, it would probably not have the following amongst cosmologists that it does today. Arguably the most impressive success of the inflationary paradigm is its providing a mechanism for understanding the origin of subtle features of the CMB. The setting for this success is a quantum treatment of the (putatively classical) story sketched in Section~\ref{field}; and it to this connection between the inflaton and the CMB that we now turn. 

The basic observable associated with the CMB is the intensity of radiation as a function of frequency and direction in the sky (Hu and Dodelson 2002). To a very good approximation, the CMB has a black-body spectrum with an average temperature of ${\bar {T}} \sim 2.73$ K ($\sim -270.42$ degrees C). But the temperature is not exactly uniform across the sky; there are small fluctuations about this mean on the order of $10^{-5}$ in size. That is: given the temperature $T(\theta,\phi)$ for a given direction $(\theta,\phi)$ in the sky, we subtract the mean temperature $\bar T$ and define a dimensionless temperature anisotropy
\be
\frac{\Delta T}{T}(\theta,\phi) \; := \; \frac{T(\theta,\phi) - {\bar {T}}}{{\bar {T}}} \; .
\ee
It is this quantity that is about $\pm 10^{-5}$ for any direction $(\theta,\phi)$. By expressing it with spherical harmonics (the analogue for the surface of a sphere, of elementary Fourier series for functions on a real interval), we can represent how the size of these irregularities in $T$ vary with angular scale. Of course, we expect the exact variation with direction in the sky, and with angular scale, to be a matter of happenstance: so we must adopt a statistical approach. So we postulate an ensemble of possible temperature functions $T(\theta,\phi)$, equipped with a probability distribution. We then use the distribution to calculate various means (averages) and dispersions (spreads), which we compare with experiment. 

The upshot of this is that for each positive integer $l$, a number $C_l$ encodes the mean size of irregularities on an angular scale of about $180 \deg / l$.  Thus $l = 1$ corresponds to being hot in one direction and cold in the opposite direction. Indeed, there is such a contribution, thanks to the Doppler effect due to the Earth's motion relative to the CMB (with the hot direction corresponding to shorter wavelengths, i.e. motion towards the CMB). The set of the $C_l$ is called the {\em angular power spectrum}. 

Of course, this statistical approach makes various assumptions, albeit defeasible ones: such as that the temperature fluctuations are Gaussian, which means that the angular power spectrum completely determines the statistics of the fluctuations. There are also other subtleties about the power spectrum such as (i) acoustic oscillations which are also a function of angular scale, and (ii) anisotropies associated with polarization effects in the CMB: but we will set these aside  (for a discussion, see Baumann and Peiris 2009). Thus the issue at hand is how these anisotropies in temperature form, and what determines their statistics. 

Inflationary cosmology proposes that these anisotropies come from small perturbations in the inflaton field, arising from its quantum fluctuations: perturbations which can then be connected (via the appropriate transfer function) to the angular power spectrum of the CMB. It is a remarkable story, spanning a complex sequence of cosmological events: but one worth briefly describing since it represents the predominant way of thinking about the origin of CMB anisotropies. (We set aside the issue of the quantum-classical transition: cf. (4) at the end of Section \ref{intro}.) 

The proposed mechanism is that as the (homogeneous) inflaton $\phi_{0}(t)$ rolls down the inflaton potential, it acquires spatially dependent quantum fluctuations $\delta\phi(t,\vec{x})$. These fluctuations mean that inflation will end at different points of space $\vec{x}$ at different times; and this leads to fluctuations in energy density after the end of inflation, ultimately giving rise to fluctuations in CMB temperature as a function of position in the sky. 

The fluctuations in the inflaton field are computed in quantum theory. One can compute a power spectrum of scalar (inflaton) fluctuations, $P_{s}(k)$ (here expressed as a function of the Fourier wavenumber $k$), whose scale ($k$) dependence is summarized by the \emph{scalar spectral index} $n_{s}$. If $n_{s} = 1$ then the power spectrum is scale-invariant, whereas $n_{s} \neq 1$ encodes deviations from scale invariance. These perturbations also lead  to the generation of primordial gravitational waves (i.e. tensor fluctuations), for which an analogous power spectrum can be computed. It is designated by $P_{t}(k)$, for which a similar tensor spectral index $n_{t}$ can also be defined. The relative strength of tensor to scalar perturbations is an important cosmological parameter, and is measured by the tensor-to-scalar ratio: $r := P_{t}(k_{\star})/P_{s}(k_{\star})$, measured at a particular reference scale $k_{\star}$. 

From these power spectra for scalar and tensor fluctuations, one can infer angular power spectra for temperature and for polarization in the CMB; and conversely, measured power spectra in the CMB can be used to infer primordial power spectra. In particular, the scalar spectral index $n_{s}$ and the tensor-to-scalar ratio $r$ are commonly used to constrain single-field slow-roll inflationary potentials. Recent Planck data [Ade et al. (Planck Collaboration) 2015] finds that:\\
\indent (i) $n_{s} = 0.968\pm 0.006$ (i.e., primordial fluctuations are nearly scale invariant), at a 68\% confidence level;  and\\
\indent (ii) $r<0.11$, at a 95\% confidence level. 

Thus inflation---even its simpler, single-field slow-roll, models---indeed provides potentials that are consistent with the statistics of the CMB. But how many such single-field slow-roll models are consistent with CMB measurements? It is to this question, and the underlying under-determination that its answer reveals, that we now turn. 

\subsection{A plethora of models}\label{pleth}
 
The fact that inflation provides a mechanism for understanding both (i) very large-scale homogeneous features of our universe (Section \ref{idea}) and (ii) much smaller-scale inhomogeneities apparent in  the CMB (Section \ref{cause}), suggests that inflationary models should be highly constrained. Indeed, they are. But there remains a wide variety of possible inflationary models which yield these impressive successes. And this variety is not a mere matter of (a) margins of error, or (b) simplicity: like a case where a physical field, e.g. a potential function, can be ascertained either (a) only within certain bounds, or (b) only by assuming it is simple in some precise sense (e.g. being a polynomial of degree at most four). Instead, inflationary models that differ by substantially more than matters (a) and (b) yield predictions that are the same---at least, the same so far as we can confirm them. 

Agreed: one might hope that this under-determination is not a problem with the theory of inflation per se, but only a reflection of the effective nature of the theory. That is: although primordial inflation operates at very high energies ($\sim10^{16}$ GeV), it is expected to be some low-energy limit of a fundamental theory whose details are yet to be worked out (Cheung et al., 2008):  and which will, one hopes, overcome the under-determination. Be that as it may, we now give some details about the wide variety of possible inflationary models.

Broadly speaking, inflationary model building is pursued along three main avenues: in terms of:\\
\indent (i)  a single scalar field (as outlined in Section~\ref{field}); or\\
\indent (ii) multiple scalar fields (known as``multi-field inflation"); or \\
\indent (iii) non-scalar degrees of freedom. \\
As in Section \ref{cause}, we shall focus on (i), and on models where the scalar field is minimally coupled to gravity: which is presumably, the simplest means of realizing inflation. 

First, the good news. Such models predict negligible non-Gaussianities in the CMB, and are thus promising candidates for providing a description of the state of the very early universe. And some of them disagree with each other as regards some observations we may indeed be able to make. For example: one can categorize such models according to the range of field values $\Delta\phi$ through which the inflaton rolls (down the potential), between the the time when the largest scales now observed in the CMB were created, and the end of inflation. When this range 
$\Delta\phi$ is smaller than the Planck scale, $\Delta\phi<M_{\scriptsize\textrm{Pl}}$, we say there is ``small-field inflation'' ; whereas large-field inflation occurs when $\Delta\phi>M_{\scriptsize\textrm{Pl}}$. One observational difference between these two categories is the size of the primordial gravitational waves they predict: large-field inflation (but not small-field inflation) predicts gravitational waves that we may indeed be able to see in the near future. 

But there is also bad news: that is, a worrying plethora of models. In a bid to understand the various possibilities for inflationary models, Martin, Ringeval and Vennin (2014a) have catalogued and analyzed a total of 74(!) distinct inflaton potentials that have been proposed in the literature: all of them corresponding to a minimally coupled, slowly-rolling, single scalar field driving the inflationary expansion. And a more detailed Bayesian study (Martin et al., 2014b), expressly comparing such models with the Planck satellite's 2013  data about the CMB, shows that of a total of 193(!) possible models---where a ``model" now includes not just an inflaton potential but also a choice of a prior over parameters defining the potential---about $26\%$ of the models (corresponding to 15 different underlying potentials) are favored by the Planck data. A more restrictive analysis (appealing to complexity measures in the comparison of different models) reduces the total number of favored models to about $9\%$ of the models, corresponding to 9 different underlying potentials (though all of the ``plateau" variety). To sum up: although this is an impressive reduction of the total number of possibilities, there remains, nevertheless, an under-determination of the potential by the data: an under-determination that is not a mere matter of (a) margins of error, or (b) simplicity, of the kind noted at the start of this Section. And of course, we have not surveyed the other avenues, (ii) and (iii) above, through which inflation might be implemented: let alone the rival frameworks mentioned in footnote \ref{InflSceptm}.

Agreed, and as we said at the start of this Section: one can hope that this under-determination will be tamed as theories are developed that describe the universe at energy levels higher than those at which inflation putatively operates. But---as we shall discuss in Section \ref{prob}---this regime of yet higher energies may yield another, perhaps more serious, problem of under-determination: a problem relating to distance scales that far outstrip those we have observed, and indeed, those that we will ever observe.

\section{Confirming a multiverse theory}\label{prob}
\subsection{Eternal inflation begets a multiverse}\label{begets}
One of the more startling predictions of a variety of inflationary models is the existence of domains outside our observable horizon, where the fundamental constants of nature, and perhaps the effective laws of physics more generally, vary (Guth 1981; Guth and Weinberg 1983; Steinhardt 1983; Vilenkin 1983; Linde 1983, 1986a, 1986b). More specifically: such a `multiverse' arises through \emph{eternal inflation}, in which once inflation begins, it never ends.\footnote{This process should more accurately be described as \emph{future-eternal} inflation. Borde, Guth, and Vilenkin (2003) argue that inflation cannot be past-eternal; but see Aguirre (2007a) for a different interpretation of their results.} Eternal inflation is predominantly studied in two broad settings: (1) false-vacuum eternal inflation and (2) slow-roll eternal inflation.    

(1): False-vacuum eternal inflation can occur when the inflaton field $\phi$, which is initially trapped in a local minimum of the inflaton potential $V(\phi)$ (a state that is classically stable but quantum mechanically metastable---i.e., a `false' vacuum), either:\\
\indent (a) tunnels out of the local minimum to a lower minimum of energy-density, in particular to the `true vacuum', i.e. the true ground  state: (as described by Coleman and De Luccia (1980)); or \\
\indent (b) climbs, thanks to thermal fluctuations, over some barrier in $V(\phi)$, to the true vacuum. \\
The result is generically a bubble (i.e. a domain) where the field value inside approaches the value of the field at the true vacuum of the potential. If (i) the rate of tunneling is significantly less than the expansion-rate of the background spacetime, and-or (ii) the temperature is low enough, then the ``channels'' (a) and-or (b) by which the field might reach the true vacuum are frustrated. That is: inflation never ends, and the background inflating space-time becomes populated with an unbounded number of bubbles. (Cf. Guth and Weinberg (1983); and see Sekino et al. (2010) for a recent discussion of the various topological phases of false-vacuum eternal inflation that can arise in a simplified setting.)\footnote{After about 2000, this type of eternal inflation gained renewed interest in light of the idea of a string landscape, in which there exist multiple such metastable vacua (Bousso and Polchinski 2000;  Kachru et al. 2003; Susskind 2003; Freivogel et al. 2006).} 

(2): In  slow-roll eternal inflation, quantum fluctuations of the inflaton field overwhelm the classical evolution of the inflaton in such a way as to prolong the inflationary phase in some regions of space. When this is sufficiently probable, eternal inflation can again ensue; (Vilenkin 1983, Linde 1986a, 1986b; see Guth (2007) for a lucid summary, and Creminelli et al. (2008) for a recent discussion). It is striking that the self-same mechanism that gives rises to subtle features of the CMB (as discussed in Section~\ref{SEC:CMB}), can, under appropriate circumstances, give rise to a multiverse.

Though it is not clear how generic the phenomenon of eternal inflation is (Aguirre 2007a; Smeenk 2014), it remains a prediction of a wide class of inflationary models. So in this Section, we turn to difficulties about confirming cosmological theories that postulate a multiverse. Thus we will assume that the multiverse consists of many (possibly an infinite number of) domains, each of whose inhabitants (if any) cannot directly observe, or otherwise causally interact with, other domains; (though we will soon mention a possible exception to this).

There is a growing literature about these difficulties. It ranges from whether there could, after all, be {\em direct} experimental evidence for the other universes, to methodological debates about how we can possibly confirm a multiverse theory without ever having such direct evidence. 

We focus here on the methodological debates.\footnote{Direct experimental evidence for the other universes could perhaps be obtained from bubble collisions, i.e. from the observational imprints they might leave in our domain (see Aguirre and Johnson (2011) for a review). But such imprints have not yet been found (Feeney et al. 2011).} Thus for us, the main theme will be---not that it is hard to get evidence---but that what evidence we get may be greatly influenced by our location in the multiverse. In the jargon: we must beware of {\em selection effects}.  So one main, and well-known, theme is about the legitimacy of so-called {\em anthropic explanations}. In broad terms, the question is: how satisfactory is it to explain an observed fact by appealing to its absence being incompatible with the existence of an observer? 

We of course cannot take on the large literature about selection effects and anthropic explanations: (cf. e.g. Davies (1982); Barrow and Tipler (1988); Earman (1987); Rees (1997); Bostrom (2002); McMullin (1993, 2005, 2007); Moster\'in (2005)); Carr (2007); Landsman (2016)). We will simply adopt a scheme for thinking about these issues, which imposes a helpful `divide and rule' strategy on the various (often cross-cutting and confusing!) considerations (Section \ref{probsch}). This scheme is not original to us: it combines proposals by others---mainly: Aguirre, Tegmark, Hartle and Srednicki. Then in Section \ref{apply}, we will report results obtained by one of us (Azhar (2014, 2015, 2016)), applying this general scheme to toy models of a multiverse (in particular, models about the proportions of various species of dark matter). The overall effect will be to show that there can be a severe problem of under-determination of theory by data.
  
\subsection{A proposed scheme}\label{probsch}

 In this Section, we will adopt a scheme that combines two proposals due to others:\\
 \indent (i)  a proposal distinguishing three different problems one faces in extracting from a multiverse theory, definite predictions  about what we should expect to observe: namely, the measure problem, the conditionalization problem, and the typicality problem (Aguirre and Tegmark (2005) and Aguirre (2007b)); (Section \ref{3probs});\\
 \indent (ii)  a proposal due to Srednicki and Hartle (2007, 2010) to consider the confirmation of a multiverse theory in Bayesian terms (Section \ref{SH}).\\

\subsubsection{Distinguishing three problems: measure, conditionalization, and typicality}\label{3probs}

Our over-arching question is: given some theory of the multiverse, how should we extract predictions about what \emph{we} should expect to observe in our domain? Aguirre puts the question well:
\begin{quote}
Imagine that we have a candidate physical theory and set of
cosmological boundary conditions (hereafter denoted ${\cal T}$) that
predicts an ensemble of physically realized systems, each of which is
approximately homogeneous in some coordinates and can be characterized
by a set of parameters (i.e. the constants appearing in the standard
models of particle physics and cosmology; I assume here that the laws
of physics themselves retain the same form). Let us denote each such
system a ``universe'' and the ensemble a ``multiverse''.  Given that
we can observe only one of these universes, what conclusions can we
draw regarding the correctness of ${\cal T}$, and how? (Aguirre, 2007b: 368--369)
\end{quote}
Thus we want to somehow define the probability, assuming ${\cal T}$, of a given value, $\vec{p}$ say, of a set of observables. But calculational complexities and selection effects make this difficult.  Aguirre goes on to propose a helpful `divide and rule' strategy, which systematizes the various considerations one has to face. In effect, there are three  problems, which we now describe.\footnote{Features of these problems are implemented in actual calculations by, for example, Tegmark (2005) and Aguirre and Tegmark (2005). But a comprehensive treatment, with each of the three components explored in detail, has yet to be carried out.}\\

\emph{5.2.1.A: The measure problem}: To define a probability distribution $P$, we first need to specify the sample space: the type of object $M$ that are its elements---traditionally called `outcomes', with sets of them being called `events'. Each observable will then be a random variable on the sample space, so that for a  set of observables $\vec{p}$, $P(\vec{p})$ is well-defined. One might take each $M$ to be a domain in the sense above, so that $P(\vec{p})$ represents the probability of $\vec{p}$ occurring in a randomly chosen domain: where $P$ may, or may not, be uniform on the sample space.\footnote{One imagines that the domains are defined as subsets of a single eternally-inflating (classical) space-time: cf. Vanchurin (2015).}  But it is not clear a priori that domains should be taken as the outcomes, i.e. the basic alternatives. For suppose, for example, that ${\cal T}$ says domains in which our observed values occur are much smaller than domains where they do not occur.  So if we were to split each of these latter domains into domains with the size of our domain, the probabilities would radically change. In short, the problem is that there seems no a priori best way of selecting the basic outcomes.

Besides, this problem is made worse by various infinities that arise in eternal inflation. Mathematically natural measures over reasonable candidates for the outcomes often take infinite values, and so probabilities often become ill-defined; (including when they are taken as ratios of measures). Various regularization schemes have been introduced to address such issues; but the probabilities obtained are not independent of the scheme used. This predicament---the need, for eternally inflating space-times, to specify outcomes and measures, and to regularize the measures so as to unambiguously define probabilities---is known as the \emph{measure problem}. For a review of recent efforts, cf. Freivogel (2011). (Note that this measure problem is {\em not} the same as the problems besetting defining measures over initial states of a cosmological model, as in (3) at the end of Section \ref{intro}: cf. Carroll (2014: footnote 13).\\

\emph{5.2.1.B: The conditionalization problem}: Even if one has a solution to the measure problem, a second problem arises. It is expected, that for any reasonable ${\cal T}$ and any reasonable solution to the measure problem, the probabilities for what \emph{we} will see will be small. For in eternal inflation, generically,  much of the multiverse  is likely to not resemble our domain. Should we thus conclude that all models of eternal inflation are disconfirmed? We might instead propose that we should restrict attention to domains (or more generally: regions) of the multiverse, where we can exist. That is, we should {\em conditionalize}: we should excise part of the sample space and renormalize the probability distribution, and then compare the resulting distribution with our observations. This process of conditionalization can be performed in three main ways; (see Aguirre and Tegmark 2005):\\
\indent (i) we can perform no conditionalization at all---known as the ``bottom-up" approach;\\
\indent (ii) we can try to characterize our observational situation by selecting observables in our models that we believe are relevant to our existence (and hence for any future observations and experiments that we may make)---known as the ``anthropic approach";\\
\indent (iii) we can try to fix the values of each of the observables in our models, except for the observable we are aiming to predict the value of---known as the ``top-down" approach. 

As one might expect, there are deep, unresolved issues, both technical and conceptual, for both (ii) and (iii) above. For the anthropic approach, (ii), one faces the difficult question of how precisely to characterize our observational situation, and so which values of which observables we should conditionalize on. In the top-down approach, (iii), it is unclear how we can perform this type of conditionalization in a practical way.

 And on both approaches, we expect the observable we aim to predict the value of and the conditionalization scheme to be separated in an appropriate way---but it is not clear how to go about doing this. It is natural to require that the observable being predicted:\\
 \indent (a) is correlated with the conditionalization scheme (otherwise the conditionalization scheme would play no role in the predictive framework), but: \\
 \indent (b) is not perfectly correlated with the conditionalization scheme (otherwise one would be open to the charge of circularity). \\
So when it is not clear exactly how observables are correlated with the defining features of a conditionalization scheme---as indeed it is not in eternal inflation!---the need to strike a balance between these two requirements amounts to a difficult problem. In short, the problem is: how can we distinguish the observable to be predicted, from the defining features of the conditionalization scheme? (See Garriga and Vilenkin (2008, Section III) who mention such concerns, and Hartle and Hertog (2013).)\\

\emph{5.2.1.C: The typicality problem}: Even if one has solved (at least to one's own satisfaction!) both the measure and conditionalization problems, a third problem remains: the \emph{typicality problem}. Namely: for any appropriately conditionalized probability distribution, how typical should we expect our observational situation to be, amongst the domains/regions, to which the renormalized probability distribution is now restricted? In other words: how much ``away from the peak, and under the tails'' can our observations be, without our taking them to disconfirm our model?  In the next Section, we will be more precise about what we mean by `typicality'; but for now, the intuitive notion will suffice.

 Current discussions follow one of two distinct approaches. One approach asserts that we should always assume typicality with respect to an appropriately conditionalized distribution. This means we should assume Vilenkin's ``principle of mediocrity", or something akin to it (Gott (1993); Page (1996); Vilenkin (1995); Bostrom (2002); Garriga and Vilenkin (2008)). The other approach asserts that the assumption of typicality is just that---an assumption---and is thus subject to error. So one should allow for a spectrum of possible assumptions about typicality; and in aiming to confirm a model of eternal inflation, one tests the typicality assumption, in conjunction with the measure and conditionalization scheme under investigation. This idea was developed within a Bayesian context most clearly by Srednicki and Hartle (2007, 2010); and it is to a discussion of their scheme that we now turn. 
 
\subsubsection{The Srednicki-Hartle proposal: frameworks}\label{SH}

Srednicki and Hartle (2010) argue that in what they call a ``large universe'' (i.e. a multiverse), we test what they dub a \emph{framework}: that is, a \emph{conjunction} of four items: a cosmological (multiverse) model $\cal T$, a measure, a conditionalization scheme, and an assumption about typicality. (Agreed, the model $\cal T$ might specify the measure; but it can hardly be expected to specify the conditionalization scheme and typicality assumption.)

If such a framework does not correctly predict our observations, we have license to change any one of its conjuncts,\footnote{Philosophers of science will of course recognize this as an example of the Duhem-Quine thesis.} and to then compare the distribution derived from the new framework with our observations. One could, in principle, compare frameworks by comparing probabilities for our observations against one another (i.e., by comparing likelihoods); or one can formulate the issue of framework confirmation in a Bayesian setting (as Srednicki and Hartle do). 

To be more precise, let us assume that a multiverse model $\mathcal{T}$ includes a prescription for computing a measure, i.e. includes a proposed solution to the measure problem: this will simplify the notation.   So given such a   model $\mathcal{T}$, a conditionalization scheme $\mathcal{C}$, and a typicality assumption, which for the moment we will refer to abstractly as $\xi$: we aim to compute a probability distribution for the value  $D_{0}$ of some  observable. We write this as $P(D_{0}|\mathcal{T}, \mathcal{C}, \xi)$. Bayesian confirmation of frameworks then proceeds by comparing posterior distributions $P(\mathcal{T}, \mathcal{C}, \xi|D_{0})$, where
\begin{equation}\label{EQN:BayesProb}
P(\mathcal{T}, \mathcal{C}, \xi|D_{0}) = \frac{P(D_{0}|\mathcal{T}, \mathcal{C}, \xi) P(\mathcal{T}, \mathcal{C}, \xi)}{\sum_{\{\mathcal{T'}, \mathcal{C'}, \xi'\}}P(D_{0}|\mathcal{T'}, \mathcal{C'}, \xi') P(\mathcal{T'}, \mathcal{C'}, \xi')}, 
\end{equation}
and $P(\mathcal{T}, \mathcal{C}, \xi)$ is a prior over the framework $\{\mathcal{T}, \mathcal{C}, \xi\}$. 

How then do we implement typicality assumptions? Srednicki and Hartle (2010) develop a method for doing so by assuming there are a finite number $N$ of locations where our observational situation obtains. Assumptions about typicality are made through ``xerographic distributions" $\xi$, which are probability distributions encoding our beliefs about at {\em which} of these $N$ locations \emph{we} exist. So if there are space-time locations $x_{A}$, with $A= 1,2,\dots,N$, where our observational situation obtains, the xerographic distribution $\xi$ is a set of $N$ numbers $\xi\equiv\{\xi_{A}\}_{A=1}^{N}$, such that $\sum_{A=1}^{N}\xi_{A}=1$. Thus typicality is naturally thought of as the uniform distribution, i.e., $\xi_{A}= 1/N$. Assumptions about various forms of atypicality correspond to deviations from the uniform distribution. Likelihoods $P(D_{0}|\mathcal{T}, \mathcal{C}, \xi)$ in Eq.~(\ref{EQN:BayesProb}) can then be computed via: $P(D_{0}|\mathcal{T}, \mathcal{C}, \xi) = \sum_{A=1}^{N}\xi_{A}P(D_{0}[A]|\mathcal{T}, \mathcal{C})$, where $D_{0}[A]$ denotes that $D_{0}$ occurs at location $A$. (Admittedly, this equation is somewhat schematic: see Srednicki and Hartle (2010: Appendix B) and Azhar (2015:  Section II A and III) for more detailed implementations).

In this way, different assumptions about typicality, expressed as different choices of $\xi$, can be compared against one another for their predictive value. Azhar (2015) undertakes the task of explicitly comparing different assumptions about typicality  in simplified multiverse cosmological settings. He shows that for a fixed model, the assumption of typicality i.e. a uniform xerographic distribution (with respect to a particular reference class) achieves the maximum likelihood for the data $D_{0}$ considered. But typicality does {\em not} necessarily lead to the highest likelihoods for our data, if one  allows different models to compete  against each other. 

This conclusion is particularly interesting when for some model, the assumption of typicality is not the most natural assumption. Hartle and Srednicki (2007: Section II) make the point with an amusing parable. If we had a model according to which there were vastly more sentient Jovians than Earthlings, we would surely be wrong, {\em ceteris paribus}, to regard the natural corresponding framework as disconfirmed merely by the fact that according to it, we are not typical observers. That is: it is perfectly legitimate for a model to suggest a framework that describes us as untypical. And if we are presented with such a model, we should not demand that we instead accept another model under which typicality is a reasonable assumption.  Indeed, demanding this  may lead us to choose a framework that is less well confirmed, in Bayesian terms.\footnote{{\em Aficionados} of  the threat of ``Boltzmann brains" in discussions of the cosmological aspects of foundations of  thermal physics  will recognize the logic of the Jovian parable. We cannot here go into details of the analogy, and the debates about Boltzmann brains; (cf. e.g. Albrecht and Sorbo 2004; De Simone et al. 2010). Suffice it to say that the main point is: a model that implies the existence of many such brains may be implausible, or disconfirmed, or have many other defects; but it is not to be rejected, just because it implies that we---happily embodied and enjoying normal lives!---are not typical. In other words: it takes more work to rebut some story that there are zillions of  creatures who are very different from me in most respects, but nevertheless feel like me (``share my observations/experiences''), than just the thought `if it were so, how come I am not one of them?'. For the story might also contain a good account of how and why, though I feel like them, I am  otherwise so different.}

\subsection{Different frameworks, same prediction}\label{apply}

As an example of how the scheme described in Section~\ref{SH} can lead to further interesting insights, we describe recent work that investigates simplified frameworks in toy models of a multiverse. The results will again illustrate the main theme of this chapter: the under-determination of theory by data.\footnote{This section is based on Azhar (2014, 2016), which both build upon the work of Aguirre and Tegmark (2005). The scheme we use does not explicitly address the measure problem; we simply assume there is some solution, so that probability distributions over observables of interest can indeed be specified.}

Aguirre and Tegmark (2005) considered a multiverse scenario in which the total number of species of dark matter can vary from one domain to another. They assumed the occurrence of the different species to be probabilistically independent of one another; and then investigated how different conditionalization schemes can change the prediction of the total number of dominant species (where two or more species are called `dominant' when they have comparable densities, each of which is much greater than the other species' density). Azhar (2016) extended this analysis, by considering (i) probabilistically correlated species of dark matter, and (ii) how this prediction varies when one makes various assumptions about our typicality, in the context of various conditionalization schemes. We will thus conclude this Section by outlining  one example of this sort of analysis, and highlighting the conclusions about under-determination thus obtained.

In the notation of Section~\ref{3probs}; assume that $\mathcal{T}$ is some multiverse model (with an associated measure), which predicts a total of $N$ distinct species of dark matter. We assume that from this theory we can derive a joint probability distribution $P(\eta_{1}, \eta_{2}, \dots, \eta_{N}|\mathcal{T})$ over the densities of different dark matter species, where the density for species $i$, denoted by $\eta_{i}$, is given in terms of a dimensionless dark matter-to-baryon ratio $\eta_{i}:=\Omega_{i}/\Omega_{b}$. We observe the \emph{total density} of dark matter $\eta_{\textrm{obs}}:=\sum_{i=1}^{N}\eta_{i}$. In fact, according to results recently released by the \emph{Planck} collaboration,  $\eta_{\textrm{obs}}\approx 5$ (Ade et al. 2015). 

In Azhar (2016), some simple probability distributions $P(\eta_{1}, \eta_{2}, \dots, \eta_{N}|\mathcal{T})$ are postulated, from which considerations of conditionalization and typicality allow one to extract predictions about the total number of dominant species of dark matter. The conditionalization schemes studied in Azhar (2016) (and indeed in Aguirre and Tegmark 2005) include examples of the ``bottom-up", ``anthropic", and ``top-down" categories discussed in Section~\ref{3probs}.\footnote{To be more precise:  in the bottom-up and top-down cases, the total number $N$ of species is fixed by assumption, and one looks for the total number of dominant species;  while in the anthropic case, $N$ is allowed to vary, and one looks for the total number $N$ of equally contributing components.} And considerations of typicality are implemented in a straightforward way, by taking typical parameter values to be those that correspond to the peak of the (appropriately conditionalized) probability distribution.\footnote{The issue of how this characterization of typicality relates to xerographic distributions is subtle. It should be clear that for a finite number of domains, the assumption of typicality as encoded by a uniform xerographic distribution corresponds to assuming that we (and thus the parameter values we observe) are randomly `selected' from among the domains, and this \emph{is} typicality as understood in the straightforward way. The corresponding relationship for the case of atypicality is more nuanced. For with non-uniform xerographic distributions, one has the freedom to choose which of the domains receive which precise (xerographic) weight: a feature that is lost when one assumes that atypicality corresponds to an appropriate deviation from the peak of a probability distribution.}

Here is one example of the rudiments of this construction. Top-down conditionalization corresponds to analyzing probability distributions $P(\eta_{1}, \eta_{2}, \dots, \eta_{N}|\mathcal{T})$ along the constraint surface defined by our data, i.e., $\eta:=\sum_{i=1}^{N}\eta_{i}\approx 5$. The number of species of dark matter that contribute significantly to the total dark matter density, under the assumption of typicality, is then the number of significant components that lie under the peak of this distribution on the constraint surface. 

Azhar (2016) shows that \emph{different} frameworks can lead to precisely the same prediction. In particular, the prediction that dark matter consists of a single dominant component can be achieved through various (mutually exclusive) frameworks. Besides,  the case of multiple dominant components does not fare any better. In this sense, any future observation that establishes the total number of dominant species of dark matter will under-determine which framework could have given rise to the observation. 

The moral of this analysis is thus that if in more realistic cosmological settings, this under-determination is robust to the choice of which observable we aim to predict the value of (as one would expect), then we must accept that our observations will simply not be able to confirm any single framework for the inflationary multiverse.

\section{Envoi}\label{concl}
So much by way of surveying how cosmology, especially primordial cosmology, bears on scientific realism. Let us end by very briefly summarizing  our position. We have espoused scientific realism as a modest thesis of cognitive optimism: that we can know about the unobservable, and that indeed we do know a lot about it. Cosmology causes no special trouble for this thesis: though there are systematic limitations, even in the form of theorems, about what we can know about the global structure of spacetime (Section \ref{gen}). Besides, this thesis is well illustrated, we submit, by countless  results of modern cosmology: astonishing though these results are, as regards the vast scales of distance, time (or other quantities, such as temperature, energy and density) that they involve (Section \ref{know}). 

Of course, probing ever more extreme regimes of distance, time or these other quantities tends to call for more inventive and diverse techniques, as regards theory  as well as instrumentation. So it is unsurprising that probing the very early universe involves intractable cases of under-determination of theory by data. In the second half of the paper, we saw this in inflationary cosmology: both for ascertaining the details of the inflaton field, for example its potential (Section \ref{plan}); and for the problems of confirming a multiverse theory (Section \ref{prob}). But as we said in Section \ref{intro}, we do not see these cases of under-determination as threatening scientific realism. For it claims only that we can know about the unobservable, and indeed do know a lot about it---not that all the unobservable is knowable. \\ \\

{\em Acknowledgements}:--- FA's work is supported by the Wittgenstein Studentship in Philosophy at Trinity College, Cambridge. FA thanks Jim Hartle, Mark Srednicki, and Dean Rickles for conversations about some of the work reported in Section 5. For comments on a previous version, we thank: Anthony Aguirre, Bernard Carr, Erik Curiel, Richard Dawid, George Ellis, Michaela Massimi, Casey McCoy, John Norton, Martin Rees, Svend Rugh, Tom Ryckman, David Sloan, Chris Smeenk and Alex Vilenkin. We also thank the editor, not least for his patience.

\section{References}\label{refs}
\begin{hangparas}{.25in}{1} 

Abbott, B. et al. (2016) ``Observation of Gravitational Waves from a Binary Black Hole Merger,''  {\em Physical Review Letters} 116: 061102.\\

Ade, P. A. R. et al.~[\emph{Planck} Collaboration] (2015) ``\emph{Planck} 2015 Results. XX. Constraints on Inflation," arXiv:1502.02114.\\

Albrecht, A. and Sorbo, L. (2004) ``Can the Universe Afford Inflation?,'' {\em Physical Review} D 70: 063528.\\ 

Alpher, R. and Herman, R. (1948) ``Evolution of the Universe,'' {\em Nature} 162: 774--5.\\

Aguirre, A. (2007a) ``Eternal Inflation, Past and Future," arXiv:0712.0571.\\

------ (2007b) ``Making Predictions in a Multiverse: Conundrums, Dangers, Coincidences," in B. Carr ed.\\

Aguirre, A. and Johnson, M. C. (2011) ``A Status Report on the Observability of Cosmic Bubble Collisions," \emph{Reports on Progress in Physics} 74: 074901.\\

Aguirre, A. and Tegmark, M. (2005) ``Multiple Universes, Cosmic Coincidences, and Other Dark Matters," \emph{Journal of Cosmology and Astroparticle Physics} 01(2005)003.\\

Azhar, F. (2014) ``Prediction and Typicality in Multiverse Cosmology," \emph{Classical and Quantum Gravity} 31: 035005.\\

------ (2015) ``Testing Typicality in Multiverse Cosmology," \emph{Physical Review} D 91: 103534.\\

------ (2016) ``Spectra of Conditionalization and Typicality in the Multiverse," \emph{Physical Review} D 93: 043506.\\

Barrow, J. (2011) {\em The Book of Universes}, London: Bodley Head.\\

Barrow, J. and Tipler, F. (1988) {\em The Anthropic Cosmological Principle}, Oxford:
Oxford University Press; (original edition 1986).\\

Baumann, D. and Peiris, H. V. (2009) ``Cosmological Inflation: Theory and Observations," \emph{Advanced Science Letters} 2: 105--120.\\

Beisbart, C. (2009) ``Can we Justifiably Assume the Cosmological Principle in order to Break Model Under-Determination in Cosmology?,'' {\em Journal of General Philosophy of Science} 40: 175--205.\\

Beisbart, C. and Jung, T. (2006) ``Privileged, Typical or Not Even That? Our Place in the World According to the Copernican and Cosmological Principles,''  {\em Journal of General Philosophy of Science} 37: 225--256.\\

Borde, A., Guth, A. H. and Vilenkin, A. (2003) ``Inflationary Spacetimes Are Incomplete in Past Directions," \emph{Physical Review Letters} 90: 151301.\\

Bostrom, N. (2002) {\em Anthropic Bias: Observation Selection Effects in Science and Philosophy}, New York: Routledge.\\

Bousso, R. and Polchinski, J. (2000) ``Quantization of Four-Form Fluxes and Dynamical Neutralization of the Cosmological Constant," 
 \emph{Journal of High Energy Physics} 06(2000)006.\\

Brandenberger, R. H. (2013) ``Unconventional Cosmology,'' in G. Calcagni, L. Papantonopoulos, G. Siopsis and N. Tsarnis (eds.) \emph{Quantum Gravity and Quantum Cosmology} (Lecture Notes in Physics, 863), Berlin: Springer; arXiv:1203.6698.\\

------ (2014) ``Do we have a Theory of Early Universe Cosmology?,''  {\em Studies in the History and Philosophy of Modern Physics} 46: 109--121.\\

------ (2016) ``Initial Conditions for Inflation - A Short Review," arXiv:1601.01918.\\
 
Butterfield, J. (2012) ``Under-determination in Cosmology: An Invitation,'' {\em The 
Aristotelian Society Supplementary Volume 2012}, 86: 1--18.\\

------ (2014) ``On Under-determination in Cosmology,'' {\em Studies in the History and Philosophy of Modern Physics} 46: 57--69. \\

------ (2014a) ``Reduction, Emergence and Renormalization,'' {\em The Journal of Philosophy} 111: 5--49. \\

Butterfield, J. and Bouatta, N. (2015) ``Renormalization for Philosophers," in T. Bigaj and C. W\"{u}thrich (eds.) {\em Metaphysics in Contemporary Physics} (Pozna\'{n} Studies in the Philosophy of the Sciences and the Humanities, vol.~104) Rodopi: 437--485. Available at: arXiv:1406.4532; \nolinkurl{http://philsci-archive.pitt.edu/10763/}\\

Butterfield, J. and Isham, C. (2001) ``Spacetime and the Philosophical Challenge of Quantum Gravity,'' in C. Callender and N. Huggett (eds.) {\em Physics meets Philosophy at the Planck Scale}, Cambridge: Cambridge University Press: 33--89. Available at: arXiv:gr-qc/9903072; \nolinkurl{http://philsci-archive.pitt.edu/1915/}\\

Ca\~{n}ate, P., Pearle, P. and Sudarsky, D. (2013) ``Continuous Spontaneous Localization Wave Function Collapse Model as a Mechanism for the Emergence of Cosmological Asymmetries in Inflation," {\em Physical Review} D 87: 104024.\\

Carr, B. (ed.) (2007) {\em Universe or Multiverse?}, Cambridge: Cambridge University Press.\\ 

------ (2014) ``Metacosmology and the Limits of Science," in M. Eckstein, M. Heller, S. Szybka (eds.) {\em Mathematical Structures of the Universe}, Krak\'{o}w: Copernicus Center Press.\\

Carroll, S. (2014) ``In What Sense is the Early Universe Fine-Tuned?," forthcoming in B. Loewer, E. Winsberg, and B. Weslake (eds.) {\em TimeÕs Arrows and the Probability Structure of the World},  Cambridge MA: Harvard University Press; arXiv:1406.3057.\\

Chakravartty, A. (2011) ``Scientific Realism,'' in E. N. Zalta (ed.) {\em The Stanford Encyclopedia of Philosophy}, online at: \nolinkurl{http://plato.stanford.edu/entries/scientific-realism/}\\

Chamcham, K., Barrow, J., Saunders, S. and Silk, J. (eds.) (2017) {\em The Philosophy  of Cosmology}, forthcoming: Cambridge: Cambridge University Press.\\

Chandrasekhar, S. (1939) {\em An Introduction to the Study of Stellar Structure}, Chicago: University of Chicago Press (Dover reprint: 1957).\\

Cheung, C., Fitzpatrick, A. L., Kaplan, J. and Senatore, L. (2008) ``The Effective Field Theory of Inflation,"  \emph{Journal of High Energy Physics} 03(2008)014.\\

Clarkson, C. and Maartens, R. (2010) ``Inhomogeneity and the Foundations of Concordance Cosmology,'' {\em Classical and Quantum Gravity} 27: 124008. \\

Cleland, C. (2002) ``Methodological and Epistemic Differences between Historical Science and Experimental Science,'' {\em Philosophy of Science}, 69: 447--451.\\

Coleman, S. and De Luccia, F. (1980) ``Gravitational Effects on and of Vacuum Decay," \emph{Physical Review} D 21: 3305--3315.\\

Colin, S. and Valentini, A. (2016) ``Robust Predictions for the Large-scale Cosmological Power Deficit from Primordial Quantum Nonequilibrium," \emph{International Journal of Modern Physics} D 25: 1650068; arXiv:1510.03508v3.\\

Creminelli, P., Dubovsky, S., Nicolis, A., Senatore, L. and Zaldarriaga, M. (2008) ``The Phase Transition to Eternal Inflation," \emph{Journal of High Energy Physics} 09(2008)036.\\

Curiel, E. (1999) ``The Analysis of Singular Spacetimes," {\em Philosophy of Science}, 66: Supplement S119--S145. Proceedings of the 1998 Biennial Meetings of the Philosophy of Science Association. Part I: Contributed Papers.Ê Stable URL: \nolinkurl{http://www.jstor.org/stable/188766}.Ê A more recent, corrected, revised, and extended version of the published paper is available at: \nolinkurl{http://strangebeautiful.com/phil-phys.html}.\\

------ (2015) ``Measure, Topology and Probabilistic Reasoning in Cosmology," arXiv: 1509.01878.\\

Davies, P. (1982) {\em The Accidental Universe}, Cambridge: Cambridge University Press.\\

Dawid, R. (this volume) ``Scientific Realism and High-Energy Physics''.\\

------  (2013)  {\em String Theory and the Scientific Method}, Cambridge: Cambridge University Press.\\

De Simone, A., Guth, A. H., Linde, A.,  et al. (2010) ``Boltzmann Brains and the Scale-Factor Cutoff Measure of the Multiverse,'' {\em Physical Review} D 82: 063520.\\

Durrer, R. (2015) ``The Cosmic Microwave Background: The History of its Experimental Investigation and its Significance for Cosmology,'' {\em Classical and Quantum Gravity} 32: 124007.\\

Earman, J. (1987) ``The SAP Also Rises: A Critical Examination of the Anthropic Principle,'' {\em American Philosophical Quarterly} 24: 307--317.\\

------ (1995) {\em Bangs, Crunches, Whimpers and Shrieks}, Oxford: Oxford University Press.\\

Earman, J. and Mosterin, J. (1999) ``A Critical Look at Inflationary Cosmology,'' {\em Philosophy of Science} 66: 1--49.\\

East, W. E., Kleban, M., Linde, A. and Senatore, L. (2015) ``Beginning Inflation in an Inhomogeneous Universe," arXiv:1511.05143.\\

Ellis, G. F. R. (1975) ``Cosmology and Verifiability,"  {\em Quarterly Journal of the Royal Astronomical Society} 16: 245--264.\\

------ (1999) ``Before the Beginning: Emerging Questions and Uncertainties,'' {\em Astrophysics and Space Science} 269--270: 693--720.\\

------ (1999a) ``83 Years in General Relativity and Cosmology: Progress and Problems,'' {\em Classical and Quantum Gravity} 16: A37--A75.\\

------ (2007) ``Issues in Philosophy of Cosmology,'' in Part B of
J. Butterfield and J. Earman (eds.) {\em Philosophy of Physics}, Elsevier, volume 2 of the North Holland series, {\em The Handbook of Philosophy of Science}: 1183--1286; arXiv:astro-ph/0602280.\\

------ (2014) ``On the Philosophy of Cosmology," {\em Studies in the History and Philosophy of Modern Physics} 46: 5--23.\\

------ (2016) ``Cosmology, Cosmologia and the Testing of Cosmological Theories," forthcoming in K. Chamcham et al. (eds.).\\

Ellis, G. F. R., Nel, S. D., Maartens, R., Stoeger, W. R. and Whitman, A. P. (1985) ``Ideal Observational Cosmology,'' {\em Physics Reports} 124: 315--417.\\

Ellis, G. and Silk, J. (2014) ``Defend the Integrity of Physics," {\em Nature} 516: 321--323.\\

Freivogel, B. (2011) ``Making Predictions in the Multiverse," \emph{Classical and Quantum Gravity} 28: 204007.\\

Freivogel, B., Kleban, M., Mart\'{i}nez, M. R. and Susskind, L. (2006) ``Observational Consequences of a Landscape," \emph{Journal of High Energy Physics} 03(2006)039.\\

Garriga, J. and Vilenkin, A. (2008) ``Prediction and Explanation in the Multiverse," \emph{Physical Review} D 77: 043526.\\

Giere, R. (1999) {\em Science Without Laws}, Chicago: University of Chicago Press.\\

Glymour, C. (1977) ``Indistinguishable Spacetimes and the Fundamental Group" in  J. Earman, C. Glymour and J. Stachel (eds.) {\em Foundations of Spacetime Theories}, Minnesota Studies in Philosophy of Science vol. 8, University of Minnesota Press: 50--60.\\

Gott III, J. R. (1993) ``Implications of the Copernican Principle for our Future Prospects," \emph{Nature} 363: 315--319.\\

Guth, A. H. (1981) ``Inflationary Universe: A Possible Solution to the Horizon and Flatness Problems," \emph{Physical Review} D 23: 347--356.\\

------ (1997) {\em The Inflationary Universe}, London: Penguin.\\

------ (2007) ``Eternal Inflation and its Implications," \emph{Journal of Physics A: Mathematical and Theoretical} 40: 6811--6826.\\

Guth, A. H., Kaiser, D. and Nomura, Y. (2014) ``Inflationary Paradigm after Planck 2013,'' \emph{Physics Letters} B 733: 112--119.\\

Guth, A. H. and Weinberg, E. J. (1983) ``Could the Universe Have Recovered From a Slow First-Order Phase Transition?," \emph{Nuclear Physics} B 212: 321--364.\\

Hacking, I. (1983) {\em Representing and Intervening: Introductory Topics in the Philosophy of Natural Science},  Cambridge: Cambridge University Press.\\

Hartle, J. and Hertog, T. (2013) ``Anthropic Bounds on $\Lambda$ From the No-Boundary Quantum State," \emph{Physical Review} D 88: 123516.\\

Hartle, J.B. and Srednicki, M. (2007) ``Are We Typical?," \emph{Physical Review} D 75: 123523.\\

Hollands, S. and Wald, R. (2002) ``Essay: An Alternative to Inflation," {\em General Relativity and Gravitation} 34: 2043--2055.\\

Hu, W. and Dodelson, S. (2002) ``Cosmic Microwave Background Anisotropies," \emph{Annual Review of Astronomy and Astrophysics} 40: 171--216.\\

Ijjas, A., Steinhardt, P. J. and Loeb, A. (2013) ``Inflationary Paradigm in Trouble After Planck2013," \emph{Physics Letters} B 723: 261--266.\\

Kachru, S., Kallosh, R., Linde, A. and Trivedi, S. P. (2003) ``de Sitter Vacua in String Theory," \emph{Physical Review} D 68: 046005.\\

Kaler, J. (2006) {\em The Cambridge Encyclopedia of the Stars}, Cambridge: Cambridge University Press.\\

Kleban, M. and Senatore, L. (2016) ``Inhomogeneous Anisotropic Cosmology," arXiv: 1602.03520.\\

Kolb, E. and Turner, M. (1990) {\em The Early Universe}, Westview Press: Frontiers in Physics.\\

Koperski, J. (2005) ``Should we Care about Fine-Tuning,'' {\em British Journal for the Philosophy of Science}  56: 303--19.\\
 
Kragh, H. (1996)  {\em Cosmology and Controversy}, Princeton: Princeton University Press.\\

Landsman, K. (2016) ``The Fine-Tuning Argument,'' forthcoming in K. Landsman and E. van Wolde (eds.) \emph{The Challenge of Chance: A Multidisciplinary Approach from Science and the Humanities}, Berlin: Springer; arXiv:1505.05359.\\
 
Lawrie, I. (1990) {\em A Unified Grand Tour of Theoretical Physics}, Bristol: Institute of Physics Publishing, Adam Hilger.\\

Liddle, A. (2003) {\em An Introduction to Modern Cosmology}, London: John Wiley; second edition.\\

Linde, A. D. (1983) ``Chaotic Inflation," \emph{Physics Letters} 129B: 177--181.\\
 
------ (1986a) ``Eternal Chaotic Inflation," \emph{Modern Physics Letters} A 01: 81--85.\\

------ (1986b) ``Eternally Existing Self-Reproducing Chaotic Inflationary Universe," \emph{Physics Letters} B 175: 395--400.\\

------ (1990) {\em Particle Physics and Inflationary Cosmology}, New York: Harwood Academic.\\
 
Longair, M. (2003) {\em Theoretical Concepts in Physics}, Cambridge: Cambridge University Press; 
second edition (first edition 1984).\\

------ (2006) {\em Cosmic Century}, Cambridge: Cambridge University Press.\\

Maartens, R. (2011) ``Is the Universe Homogeneous?," {\em Philosophical Transactions of the Royal Society A} 369: 5115--5137 \\

Malament, D. (1977) ``Observationally Indistinguishable Spacetimes" in  J. Earman, C. Glymour and J. Stachel (eds.) {\em Foundations of Spacetime Theories}, Minnesota Studies in Philosophy of Science vol. 8, University of Minnesota Press: 61--80.\\

Manchak, J. (2009) `Can we know the Global Structure of Spacetime?', {\em Studies in History and Philosophy of Modern Physics} 40: 53--56.\\

------ (2011) `What is a Physically Reasonable Spacetime?', {\em Philosophy of Science} 78: 410--420.\\

Martin, J., Ringeval, C. and Vennin, V. (2014a) ``Encyclop\ae dia Inflationaris," \emph{Physics of the Dark Universe} 5--6: 75--235.\\
 
Martin, J., Ringeval, C., Trotta, R. and Vennin, V. (2014b) ``The Best Inflationary Models after Planck," \emph{Journal of Cosmology and Astroparticle Physics} 03(2014)039.\\

Massimi, M. and Peacock, J. (2015) ``What are Dark Matter and Dark Energy?,'' in M. Massimi et al. {\em Philosophy and the Sciences for Everyone}, London: Routledge: 32--51.\\

Massimi, M. and Peacock, J. (2015a) ``The Origins of our Universe: Laws, Testability and Observability in Cosmology,'' in M. Massimi et al. {\em Philosophy and the Sciences for Everyone}, London: Routledge: 14--32.\\

McCoy, C. D. (2015) ``Does Inflation Solve the Hot Big Bang ModelÕs Fine-Tuning Problems?," {\em Studies in the History and Philosophy of Modern Physics} 51: 23--36. \\

McMullin, E. (1993) ``Indifference Principle and Anthropic Principle in Cosmology,'' {\em Studies in the History and Philosophy of Science} 24: 359--389.\\

------ (2005) ``Anthropic Explanation in Cosmology,'' {\em Faith and Philosophy} 22: 601--614.\\

------ (2007) ``Tuning Fine-Tuning,'' in J. Barrow et al. (eds.) {\em Fitness of the Cosmos for Life: Biochemistry and Fine-Tuning}, Cambridge: Cambridge University Press: 70--94.\\

Miller, B. (2016) ``What is Hacking's Argument for Entity Realism?," {\em Synthese} 193: 991--1006.\\

Misner, C., Thorne, K. and Wheeler, J. (1973) {\em Gravitation}, San Francisco: W.H. Freeman.\\

Norton, J. (2010) ``Cosmic Confusions: Not Supporting versus Supporting Not," {\em Philosophy of Science} 77: 501--523. Available at: \nolinkurl{http://www.pitt.edu/~jdnorton/jdnorton.html} \\

------ (2011) ``Observationally Indistinguishable Spacetimes: A Challenge for any Inductivist,'' in G. Morgan (ed.) {\em Philosophy of Science Matters: The
Philosophy of Peter Achinstein}, Oxford: Oxford University Press: 164--176. Available at: \nolinkurl{http://www.pitt.edu/~jdnorton/jdnorton.html} \\

Page, D. N. (1996) ``Sensible Quantum Mechanics: Are Probabilities only in the Mind?," \emph{International Journal of Modern Physics} D 05: 583--596.\\
 
Pasaschoff, J. (1994)  {\em The Farthest Things in the Universe}, Cambridge: Cambridge University Press.\\

Penrose, R. (2004) {\em The Road to Reality}, London: Jonathan Cape.\\

Perez, A., Sahlmann, H. and Sudarsky, D. (2006) ``On the Quantum Origin of the Seeds of Cosmic Structure," {\em Classical and Quantum Gravity} 23: 2317--2354.\\

Psillos, S.  (1999) {\em Scientific Realism: How Science Tracks Truth}, London: Routledge.\\

------ (2009) {\em Knowing the Structure of Nature}, London: Palgrave Macmillan.\\

Putnam, H. (1962) ``The Analytic and the Synthetic,'' in H. Feigl and G. Maxwell (eds.) {\em Minnesota Studies in the Philosophy of Science} vol. III, Minneapolis: University of Minnesota Press 358--397.\\

Quine, W. (1953) ``Two Dogmas of Empiricism,'' in his {\em From a Logical Point of View}, Cambridge MA: Harvard University Press.\\

Rees, M. (1997) {\em Before the Beginning}, New York: Simon and Schuster.\\

------ (2003) ``Our Complex Cosmos and its Future,'' in G. Gibbons, E. Shellard and S. Rankin (eds.) {\em The Future of Theoretical Physics and Cosmology}, Cambridge: Cambridge University Press: 17--37.\\

 Rickles, D. (2008) ``Quantum Gravity: A Primer for Philosophers,'' in D. Rickles (ed.) {\em The Ashgate Companion to Contemporary Philosophy of Physics},  Aldershot: Ashgate: 262--382.\\
 
 Rovelli, C. (2007) ``Quantum Gravity,'' in J. Butterfield and J. Earman  (eds.) {\em The Handbook of  
  Philosophy of Physics}, Part B: Amsterdam: North Holland: 1287--1330.\\

Rowan-Robinson, M. (1999) {\em The Nine Numbers of the Cosmos}, Oxford: Oxford University Press.\\

------ (2011) {\em Cosmology}, Oxford: Oxford University Press: fourth edition.\\

Rugh, S. and Zinkernagel, H. (2009) ``On the Physical Basis of Cosmic Time," {\em Studies in the History and Philosophy of Modern Physics} 40: 1--19. \\

------ (2016) ``Limits of Time in Cosmology," forthcoming in K. Chamcham et al. (eds.); arXiv:1603.05449.\\

Ruiz-Lapuente, P.  (ed.) (2010) {\em Dark Energy: Observational and Theoretical Approaches}, Cambridge: Cambridge University Press.\\

Ryan, S. and Norton, A. (2010) {\em Stellar Evolution and Nucleosynthesis}, Cambridge: Cambridge University Press.\\

Schiffrin, J. S. and Wald, R. W. (2012) ``Measure and Probability in Cosmology," {\em Physical Review} D 86: 023521.\\

Sciama, D. (1971) {\em Modern Cosmology}, Cambridge: Cambridge University Press.\\

Sekino, Y., Shenker, S. and Susskind, L. (2010) ``Topological Phases of Eternal Inflation," \emph{Physical Review} D 81: 123515.\\

Shuryak, E. (2005) ``What RHIC experiments and theory tell us about properties of quark-gluon plasma?,''
 {\it Nuclear Physics} A 750: 64-83.\\

Silk, J. (1989)  {\em The Big Bang}, San Francisco: W.H. Freeman; revised and updated edition.\\

------ (2006)  {\em The Infinite Cosmos}: Oxford: Oxford University Press.\\

Singh, S. (2004) {\em Big Bang}, London: Fourth Estate.\\

Sklar, L. (1975) ``Methodological Conservatism," {\em Philosophical Review} 84: 374-400.\\

Smeenk, C. (2013) ``Philosophy of cosmology,'' in R. Batterman (ed.) {\em The Oxford Handbook of Philosophy of Physics}, Oxford: Oxford University Press: 607--652.\\

------ (2014) ``Predictability Crisis in Early Universe Cosmology," \emph{Studies in History and Philosophy of Modern Physics} 46: 122--133.\\

------ (2016) ``Testing inflation," forthcoming in K. Chamcham et al. (eds.).\\

Smoot, G. (1993) {\em Wrinkles in Time: The Imprint of Creation}; (written with Keay Davidson), New York: Little, Brown and Company.\\

Srednicki, M. and Hartle, J. (2010) ``Science in a Very Large Universe," \emph{Physical Review} D 81: 123524.\\

Stanford, P. K. (2006) {\em Exceeding our Grasp: Science, History, and the Problem of Unconceived Alternatives}, New York: Oxford University Press.\\

Stein, H. (1992) ``Was Carnap Entirely Wrong, After All?," {\em Synthese} 93: 275--295.\\

------ (1994) ``Some Reflections on the Structure of our Knowledge in Physics,"  in D. Prawitz, B. Skyrms and D. Westerst\r{a}hl (eds.) {\em Logic, Methodology and Philosophy of Science}, New York: Elsevier Science B.V.: 633--655; (Proceedings of the Ninth International Congress of Logic, Methodology and Philosophy of Science). \\

Steinhardt, P. J. (1983) ``Natural Inflation," in G. W. Gibbons, S. W. Hawking and S. T. C. Siklos (eds.) \emph{The Very Early Universe, Proceedings of the Nuffield Workshop, Cambridge, 21 June to 9 July, 1982}, Cambridge: Cambridge University Press.\\

------ (2011) ``The inflation debate,'' {\em Scientific American} 304, April 2011: 38--43.\\

Sudarsky, D. (2011) ``Shortcomings in the Understanding of Why Cosmological Perturbations Look Classical," {\em International Journal of Modern Physics} D 20: 509--552.\\

Susskind, L. (2003) ``The Anthropic Landscape of String Theory," in B. Carr ed. arXiv:hep-th/0302219.\\

Tegmark, M. (2005) ``What Does Inflation Really Predict?," \emph{Journal of Cosmology and Astroparticle Physics} 04(2005)001.\\

van Fraassen, B. (1980) {\em The Scientific Image}, Oxford: Clarendon Press.\\

Vanchurin, V. (2015) ``Continuum of Discrete Trajectories in Eternal Inflation," \emph{Physical Review} D 91: 023511.\\

Vilenkin, A. (1983) ``Birth of Inflationary Universes," \emph{Physical Review} D 27: 2848--2855.\\

------ (1995) ``Predictions from Quantum Cosmology," \emph{Physical Review Letters} 74: 846--849.\\

Wall, A. (2013) ``The Generalized Second Law Implies a Quantum Singularity Theorem," {\em Classical and Quantum Gravity} 30: 165003.\\

Wald, R. M. (1984) {\em General Relativity}, Chicago: University of Chicago Press.\\ 

Weinberg, S. (1972) {\em Gravitation and Cosmology}, New York: John Wiley.\\
  
------ (1977) {\em The First Three Minutes}, New York: Basic Books: updated 1993.\\

------ (2008) {\em Cosmology}, Oxford: Oxford University Press.\\

Woodward, J. (2003) {\em Making Things Happen: A Theory of Causal Explanation}, Oxford: Oxford University Press.

\end{hangparas}
\end{document}